\documentclass[twocolumn,pra,showpacs,final,superscriptaddress,10pt,aps,floatfix]{revtex4-1}
\usepackage[usenames]{color}
\usepackage{amsmath}
\usepackage{amssymb}
\usepackage{mathrsfs}
\usepackage{rotating}
\usepackage{bbold}
\usepackage{euscript}
\usepackage{graphicx}
\usepackage{graphics}
\usepackage{bm}
\begin{document}

\title{Valence excitation of NO$_2$ by impulsive stimulated x-ray Raman scattering}

\author{Daniel  J. Haxton}
\affiliation{Department of Physics, University of California, Berkeley CA 94720}


\begin{abstract}


The global optimum for
valence population transfer in the NO$_2$ molecule
driven by impulsive x-ray stimulated Raman scattering 
of one-femtosecond x-ray pulses tuned below the Oxygen K-edge is determined
with the Multiconfiguration Time-Dependent Hartree-Fock method,
a fully-correlated first-principles treatment that allows for the 
ionization of every electron in the molecule.
Final valence state populations 
computed in the fixed-nuclei, nonrelativistic approximation 
are reported as a function of central wavelength and intensity.
The convergence of the calculations with respect to their adjustable parameters is fully tested.
Fixing the 1fs duration but varying the central frequency and intensity of the pulse, without chirp,
orientation-averaged maximum population transfer of 0.7\% to the valence B$_1$ state is obtained at an intensity 
of 3.16$\times$10$^{17}$ W cm$^{-2}$, 
with the central frequency substantially 6eV red-detuned from the 2nd order optimum; 2.39\% is obtained
at one specific orientation.
The behavior near the global optimum, below the Oxygen K-edge, is consistent with the mechanism
of nonresonant Raman transitions driven by the near-edge fine structure oscillator strength.



\end{abstract}

\maketitle

\section{Introduction}

Recently there have been proposals for using multiple attosecond x-ray pulses to study the quantum dynamics of
valence electronic excitations in molecules~\cite{biggs_two-dimensional_2012, mukamel2013}.
These techniques, called multidimensional  attosecond x-ray spectroscopy,
promise to provide an unparalleled, time-resolved picture of evolving electronic and internuclear nuclear structures.
Pump-probe experiments using ultrafast pulses of x-ray light 
have already provided a time-resolved picture of the evolving electronic and nuclear dynamics
in several important systems [CITE GESSNER].  The amount of information
that the proposed, more sophisticated multidimensional spectroscopic methods
will provide is much greater.  
The multidimensional methods offer the chance to observe the
quantum dynamics of multiple degrees of freedom, 
extracting the correlations among
electrons and nuclei that comprise much of atomic, molecular, and chemical physics, 
observing and manipulating electronic coherences in the molecule. 
For instance, the coupled electronic and nuclear
dynamics in chromophores [CITE] involves correlated motion of several electrons, and
there is an active effort to develop better chromophores synthetically.
Experimental methods that allow
quantum correlations and coherences to be observed and manipulated directly
are invaluable for technological advancement.

The essential benefit of x-ray wavelengths lies in their elemental and chemical specificity,
as demonstrated by recent work [CITE pi/pi*].  By creating, combining and interfering
physically localized excitons in a molecule,
which then evolve, multidimensional x-ray spectroscopies offer the promise of disentangling 
correlations in time and space.
Many of the proposals for multidimensional x-ray spectroscopy rely on the creation of a coherent
valence electronic wavepacket using nominally two-photon, stimulated x-ray Raman transitions.    
Valence excitations are most relevant to real-world applications; therefore, 
x-ray Raman methods are particularly attractive.   Furthermore, they
may be implemented without regard to the relative coherence of the multiple attosecond pulses.
The broadband x-ray pulses drive the impulsive stimulated Raman process, the
excitation via the pump frequencies and the stimulated emission of Stokes frequencies, leading
to a coherent valence electronic excitation.  The 1D- and 2D-SXRS multidimensional 
spectroscopies~\cite{biggs_two-dimensional_2012, mukamel2013} combine several such
excitations, and are capable of obtaining time- and frequency-resolved 
information about electronic structure and couplings and about
the evolving internuclear geometry.   

These proposed methods require further evolution of laser technology in order to become viable.
Considerable investment is being made in next-generation light sources, both in high-harmonic
generation and free-electron lasers, and theory may provide guidance for these investments.
Given the investment and effort required to construct these experiments, reliable, comprehensive 
and unbiased theoretical predictions may provide valuable guidance.  

Providing accurate theoretical predictions for these experiments is difficult,
however, due to the number of electrons involved in the intense-field dynamics including core excitations,
and due the inherent exponential scaling of quantum
mechanical problems with respect to system size.
There have been several sophisticated theoretical and computational 
studies of multidimensional X-ray spectroscopic and related
methods including Refs.~\cite{biggs_cysteine_2013, biggs_watching_2013, shalashilin_2016, tiger}.
However, these studies have often concentrated on electronic excitations to metastable core-excited
states and neglected the continuum above the edge, 
and often they have explicitly computed the nth-order signal
without considering large-magnitude higher-order behavior that might overwhelm lower orders even
at low intensity.  

For these reasons it is desirable to test the efficiency of impulsive stimulated x-ray Raman transitions in
polyatomic molecules
in a fully nonperturbative, first-principles calculation that accounts for all the fundamental effects including multiple
ionization, stark shifts, and highly correlated electronic dynamics.
X-ray Raman transitions involve correlation of many electrons, because the excitation of the inner, core electrons
affects the field experienced by the other electrons, and because the Auger decay of the intermediate states
involves transitions of two electrons or more.  The intense field may directly excite or ionize many electrons.
Prior treatment of strong x-ray effects like cation charge state yields 
have employed rate equation models using thousands or millions of 
electronic configurations [CITE].  In contrast, the treatment of the coherent process of stimulated electronic x-ray Raman
transitions requires a coherent, wave function description.  

%

The Multiconfiguration time-dependent Hartree-Fock (MCTDHF) method~\cite{ 
Scrinzi_MCTDHF_2004,
Kato_Kono2004,
Scrinzi_MCTDHF_2005,
Nest2005,
Cederbaum2007,
Nest2007,
Levine2008,
Kato_Kono2008,
Nest2009,
KatoYamanouchi2009,
KatoKonoH2plus,
madsen,
sato_casscf_2013,
sato_ormas_2015,
ryohto_multiresolution_2016}
 is in principle capable of calculating arbitrary
nonperturbative quantum dynamics of electrons in medium-sized molecules, with all electrons active and able
to be ionized.  The ``Hartree-Fock'' part of the name makes it a misnomer, because MCTDHF employs a systematic
expansion of the wave function in terms of a time-dependent linear combination of time-dependent Slater determinants
that converges to the exact solution in the limit of many orbitals and determinants.  
We speculate that current technology 
permits converged calculations of fixed-nuclei, electronic dynamics of
organic molecules at the limits of intensity, using the MCTDHF method, and we provide
evidence for this contention in this article.

We have applied our implementation of 
MCTDHF~\cite{prolate, restricted, sincdvr, lbnl_amo_mctdhf} to predict transfer of population to valence excited electronic
states due to stimulated X-ray transitions in the NO$_2$ molecule, in the fixed-nuclei approximation.  
This implementation includes a polyatomic representation in which the scaling in terms of three-dimensional system size (times
the maximum nuclear charge, due to the discretization required for the nuclear cusp, for explicitly-represented core electrons)
is linear times logarithmic, in other words, order $N \log{N}$.  It allows for several million Slater determinants -- a number that
may be greatly increased with current technology -- 
with an arbitrary Slater determinant space.  Flexible spaces involving different shells and excitations can be defined, and we demonstrate
the convergence of the results here with respect to the orbital and Slater determinant basis.
These MCTDHF calculations test the viability of impulsive Raman transitions for driving valence electronic state population
transfer without making any assumptions about the degree of electronic excitation, correlation, or ionization, nor about the number of photons absorbed and emitted, using a quantum mechanically coherent representation.

We show that in the NO$_2$ molecule, for impulsive stimulated x-ray Raman scattering using 1fs pulses, transitions
to the $^2$B$_1$ valence electronic state dominate.  Population transfer of 0.70\% may
be driven at 2nd order by tuning below the near-edge fine structure, as in Ref.~\cite{tiger}, without orienting the molecule,
and 2.39\% or more population can be transferred once the molecule is oriented.  However,
nonlinear effects quickly set in above 10$^{16}$ W cm$^{-2}$, and a global optimization of population transfer as
a function of intensity and central frequencies appears to be driven by 
nonresonant Raman transitions, substantially red-detuned from the near-edge fine structure, 
a mechanism not considered in Ref.~\cite{tiger}.

The results indicate that multidimensional attosecond electronic X-ray Raman spectroscopies might
in general most efficiently be
performed using pulses well red-detuned from resonant edges.  Such pulses may minimize loss through direct and sequential
ionization and make use of the coherent combination of discrete and continuum edge oscillator strength, thereby
providing the greatest potential for creating coherent valence electronic wave packets
through impulsive stimulated x-ray Raman excitation.
Strong red-detuning may provide a way to efficiently perform stimulated
Raman transitions in molecules, because it prevents an excursion of the 1s electron.  

\section{MCTDHF method and representation of NO$_2$}

Briefly, the MCTDHF method~\cite{ 
Scrinzi_MCTDHF_2004,
Kato_Kono2004,
Scrinzi_MCTDHF_2005,
Nest2005,
Cederbaum2007,
Nest2007,
Levine2008,
Kato_Kono2008,
Nest2009,
KatoYamanouchi2009,
KatoKonoH2plus,
madsen,
sato_casscf_2013,
sato_ormas_2015,
ryohto_multiresolution_2016}
solves the time-dependent Schrodinger
equation using a time-dependent linear combination of Slater determinants, with time-dependent orbitals in the Slater determinants.  
The nonlinear working equations are obtained through
application of the Lagrangian variational principle~\cite{broeck, ohta} to this wave function ansatz.

Our implementation of MCTDHF~\cite{lbnl_amo_mctdhf} 
for electrons in molecules has already been described~\cite{prolate, restricted, sincdvr}.  
The representation of orbitals, the one-electron basis, using sinc basis functions is described in Ref.~\cite{sincdvr}.
For NO$_2$, we use a grid of 55$\times$55$\times$55 ($=$166375) product sinc basis functions for the orbitals.  The spacing
between the functions is 0.2975614 bohr (about 0.56 Angstrom).  As we discuss below,
this large grid spacing leads to a substantial error in the Oxygen 1$s$ ionization energy.  For comparison, we perform a few
calculations using half grid spacing, 0.1487807 bohr, and a 111x111x111 grid of Cartesian sinc basis functions.  These finer-resolution
calculations allow us to obtain a shift that we can apply to all the coarser-grid results.  As mentioned below in the photoionization section,
this shift is 2.42eV (about 66eV).  We report energies for the coarser grid calculations, and also shifted values in parenthesis and italics,
such as ``the excitation energy of the XX to YY is ZZZ (\textit{xxxx}) eV.''

For the Slater determinant list defining the many-electron basis, we use
 full configuration interaction, 23 electrons in 15 orbitals, giving 621075 Slater determinants which are contracted
to 305760 spin-adapted linear combinations and distributed among processors.  
The mean field time step was 0.02 atomic time units (approximately one half attosecond).

Complex coordinate scaling and stretching~\cite{simon} 
is applied starting at $\pm$4 bohr in the $x$, $y$, and $z$ directions.  We perform
the smooth complex scaling transformation upon the kinetic energy and derivative operators only.  Lacking a viable
method for defining the transformed two-electron operator, we do not
transform the Coulomb operators.  
%
%
The complex ray is defined as
\begin{equation}
\begin{array}{rll}
X(x) = & x + a(x-x_0) + b \sin\left(\pi \frac{x-x_0}{x_1-x_0}\right) \\
& + c \sin^3\left(\pi \frac{x-x_0}{x_1-x_0}\right)    & (x_0 \le x \le x_1) \\
= & x + a(x+x_0) + b \sin\left(\pi \frac{x+x_0}{x_1-x_0}\right) \\
& + c \sin^3\left(\pi \frac{x+x_0}{x_1-x_0}\right)  & (-x_1 \le x \le -x_0) \\
= & x & (-x_0 \le x \le x_0)
\end{array}
\end{equation}
in which $X(x)$ defines the complex coordinate ray $X$ along which the wave function is defined
as a function of the real-valued parameter $x$ in which the operators
are represented.  Smooth scaling occurs between the scaling boundary $x_0=4a_0$ and the end of the grid $x_1\approx 8.78a_0$.
The parameters $a$, $b$, and $c$ are determined by the scaling angle
and stretching
factor (half a radian and three, respectively, such that $X'(x) = 4e^{0.5i}$ at $\pm x_1$) 
and by making the fourth derivative $X''''(x)$ continuous.  
This transformation defines the new length-gauge
dipole operators $X$, $Y$, and $Z$, and the DVR weights in each
Cartesian direction (originally uniformly $w_i=\frac{1}{\Delta}$ with $\Delta$ the sinc DVR spacing)
are modified according to $w_i \rightarrow X'(x_i) w_i$.  The first derivative operator matrix elements 
are transformed as e.g. $(d/dX)_{ij} = (d/dx)_{ij} (X'(x_i) X'(x_j))^{-1/2}$, and the kinetic energy matrix elements are
defined in the DVR approximation as
\begin{equation}
\begin{array}{ll}
\left(\frac{\partial^2}{\partial X^2}\right)_{ij} = & \frac{1}{X'(x_i) X'(x_j)} \left(\frac{\partial^2}{\partial x^2}\right)_{ij} \\
& - \delta_{ij} \frac{ 3X''(x_i)^2 - 2 X'''(x) X'(x_i)}{4X'(x_i)^4}
\end{array}
\end{equation}

Below in section~\ref{convsect} we demonstrate the convergence of the population transfer results with respect to the 
parameters of the exterior complex scaling ray.

\begin{figure}
\begin{tabular}{c}
\resizebox{0.8\columnwidth}{!}{\includegraphics*[0.6in,0.6in][5.9in,4.2in]{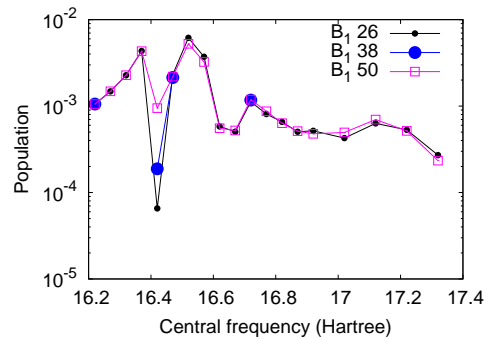}} \\
\end{tabular}
\caption{(Color online) Convergence of the $B_1$ population at 10$^{17}$ W cm$^{-2}$, with respect to the order of the Lebedev quadrature
used for the orientation average of fixed-nuclei calculations.
\label{convfig}}
\end{figure}

\section{MCTDHF calculation of impulsive Raman population transfer in NO$_2$}

\begin{figure}
\begin{tabular}{c}
\resizebox{0.8\columnwidth}{!}{\includegraphics*[0.6in,0.6in][5.9in,4.2in]{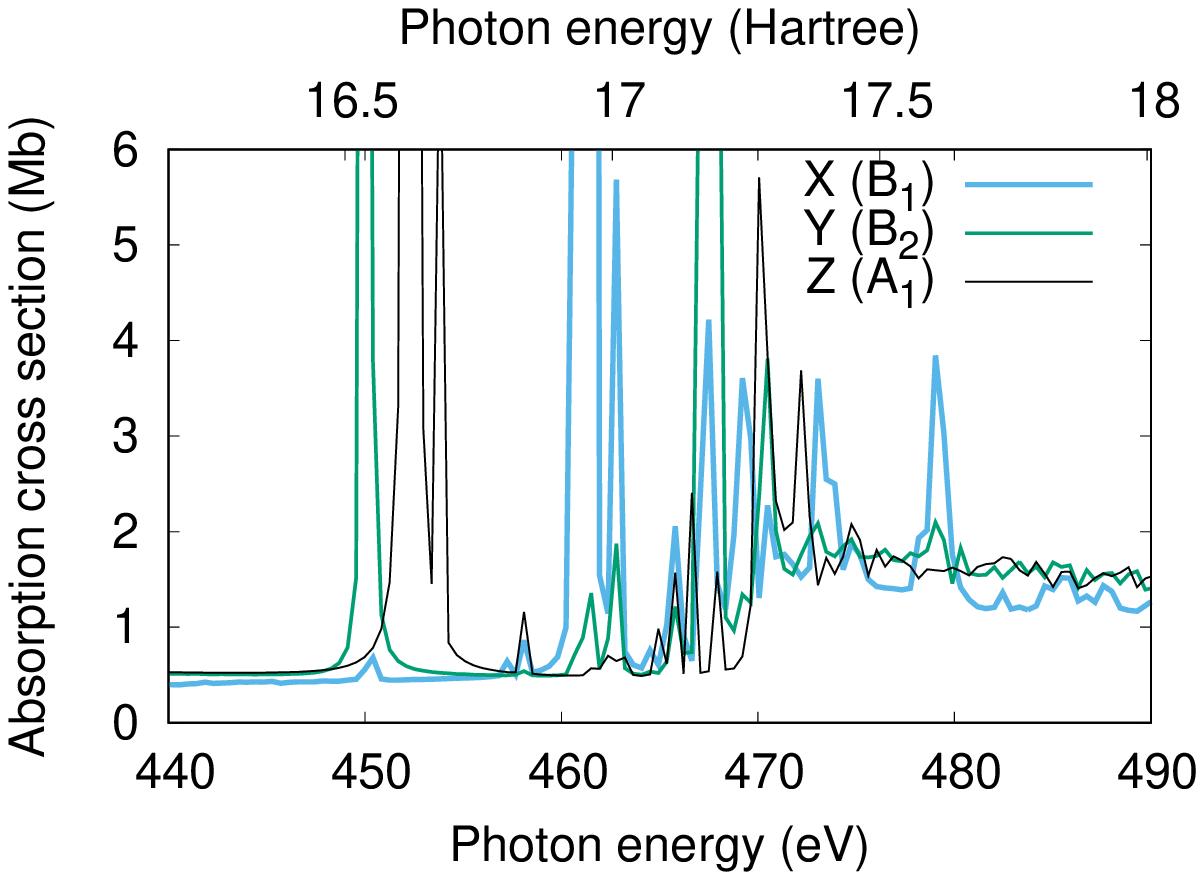}} \\
\resizebox{0.8\columnwidth}{!}{\includegraphics*[0.6in,0.6in][5.9in,4.2in]{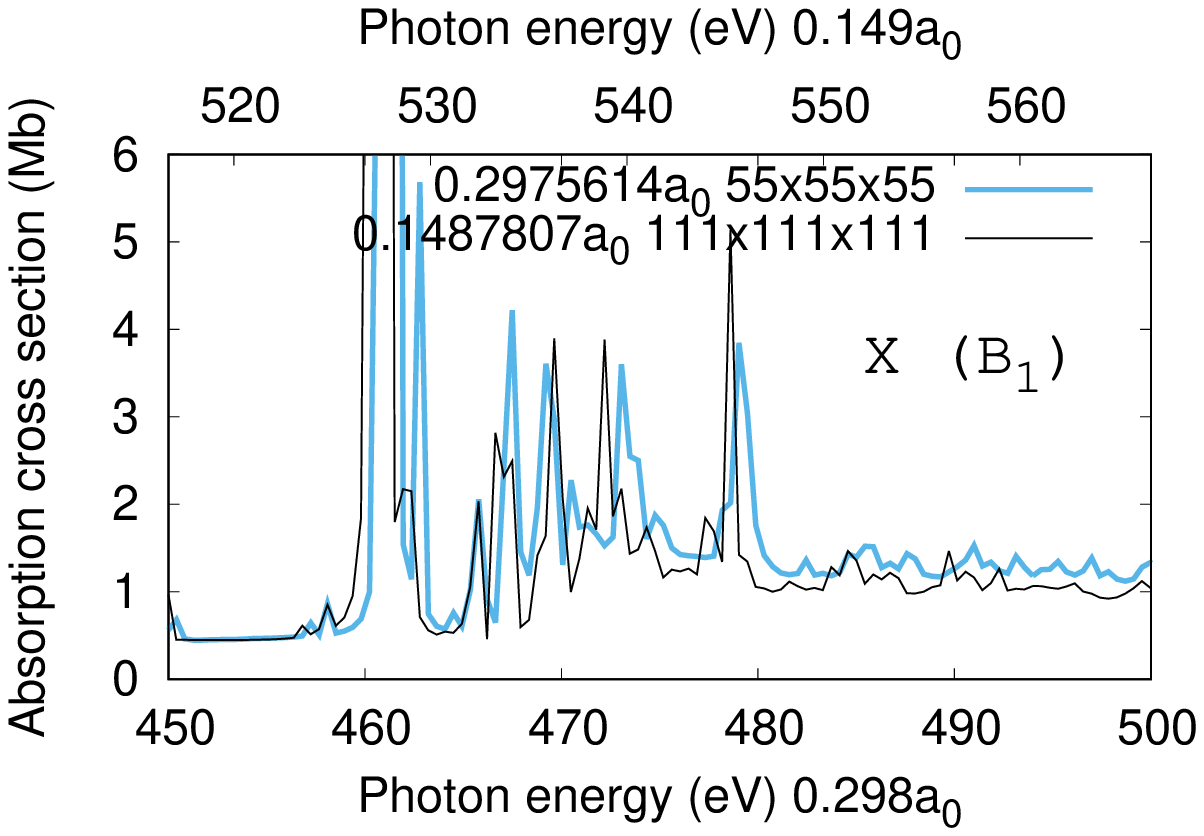}} \\
\resizebox{0.8\columnwidth}{!}{\includegraphics*[0.6in,0.6in][5.9in,4.2in]{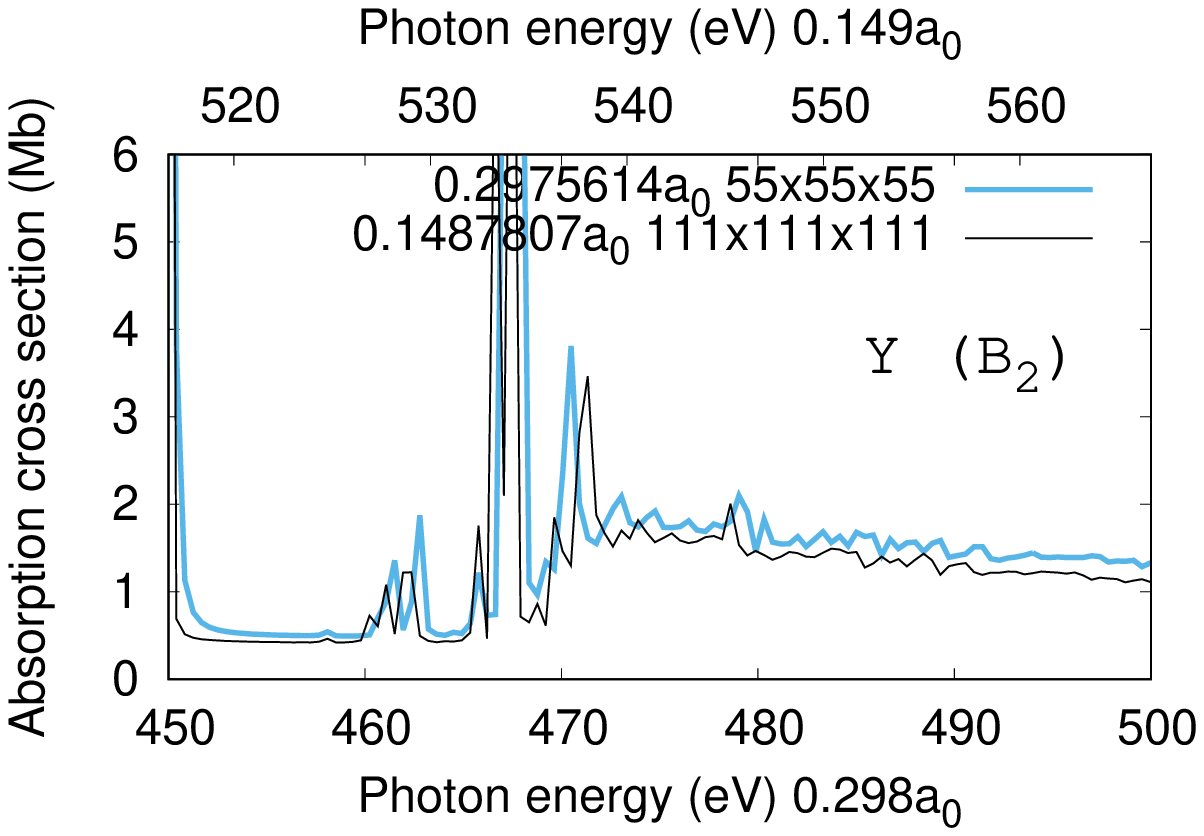}} \\
\resizebox{0.8\columnwidth}{!}{\includegraphics*[0.6in,0.6in][5.9in,4.2in]{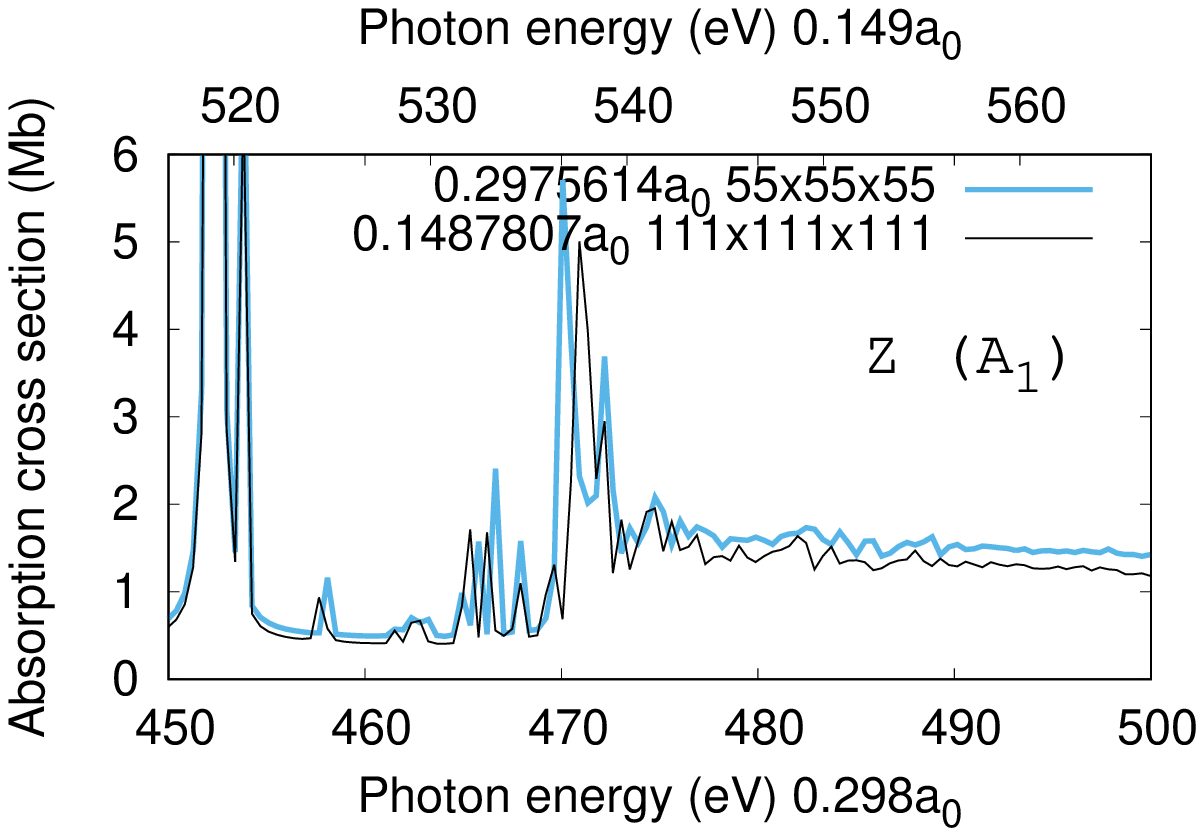}} \\
\end{tabular}
\caption{(Color online) NO$_2$ photoabsorption cross section calculated in the vicinity of the Oxygen K-edge using a 10fs MCTDHF
calculation with a weak pulse.  The top panel shows results in all three polarizations for the larger 0.2975614$a_0$ grid resolution,
55x55x55 grid
that we use for most of the calculations in this paper.  The oxygen K-edge is artificially low in these calculations.  The results with 
higher 0.1487807$a_0$ resolution, 111x111x111 grid for $x$ polarization are shown in the bottom panels, compared 
with the lower-resolution calculations.  The scales on the top and bottom for the two data sets in the bottom panels are relatively
shifted by 2.42 Hartree, approximately 66eV, which is the shift that we use when reporting the parenthetical and italicized energies
for comparison with experiment, in the body of the text.
\label{xsectfig}}
\end{figure}

We employ a pulse in the dipole approximation with central frequency $\omega$ and duration $\pi / \Omega$ as follows.
In the velocity gauge we define the vector potential
\begin{equation}
A(t) = \sin^2(\Omega t) \sin\left(\omega \left[t - \frac{\pi}{2 \Omega}\right] \right) \quad (0 \le t \le \frac{\pi}{\Omega})
\label{pulse_eqn}
\end{equation}
In the length gauge we employ the electric field $E(t) = \frac{\partial}{\partial t} A(t)$.  
For the pulses used here, with 1fs full width at half-maximum (FWHM) in time ($\pi / \Omega = $ 2fs),
the FWHM of the spectral profile, the squared Fourier transform $\vert E_z(\omega)\vert^2$
is 3.25eV.

We calculate population transfer for the valence B$_1$, B$_2$, and A$_2$ states.  We integrate the result over orientations using Lebedev
quadrature~\cite{leb1,leb2,leb3,leb4,leb5}.  
Comparing 38- and 50-point quadrature, we find that most of the results are converged with 38-point quadrature.  We have not
attempted to demonstrate the convergence of these results beyond 50-point quadrature due to computer resources.  In Figure~\ref{convfig}
we show the convergence of the $B_1$ population at 10$^{17}$ W cm$^{-2}$, with respect to the order of the Lebedev quadrature.
The convergence is good where the population transfer is large, but the sharp minimum in the population transfer at 16.4 Hartree 
requires more points for full convergence.

Due to symmetry, and
within the rotating wave approximation, only seven of the 50 points need to be calculated.  Each central frequency and intensity required approximately
10,000 cpu-hours to calculate: seven calculations, 121 processors each, and about twelve hours per calculation.

\begin{figure*}
\begin{tabular}{ccc}
\resizebox{0.66\columnwidth}{!}{\includegraphics*[0.6in,0.6in][5.9in,4.2in]{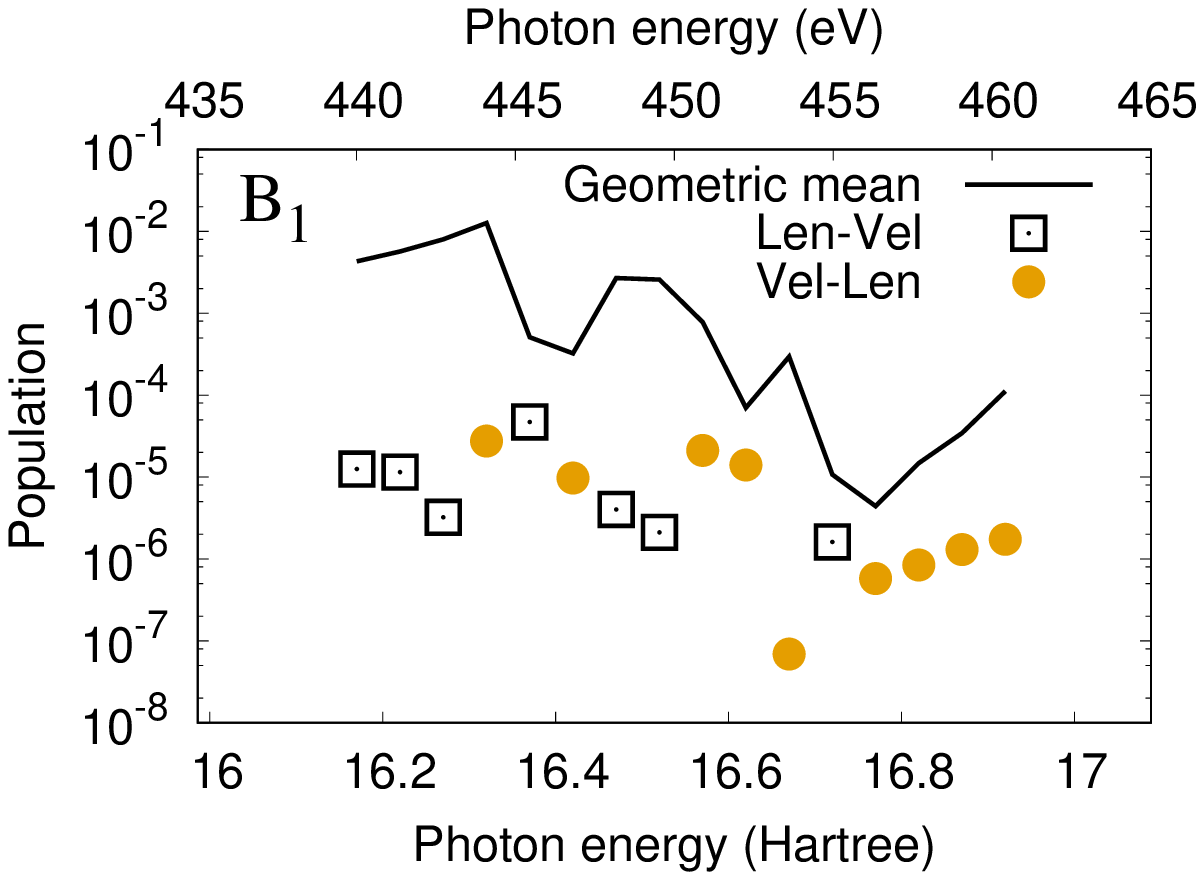}} &
\resizebox{0.66\columnwidth}{!}{\includegraphics*[0.6in,0.6in][5.9in,4.2in]{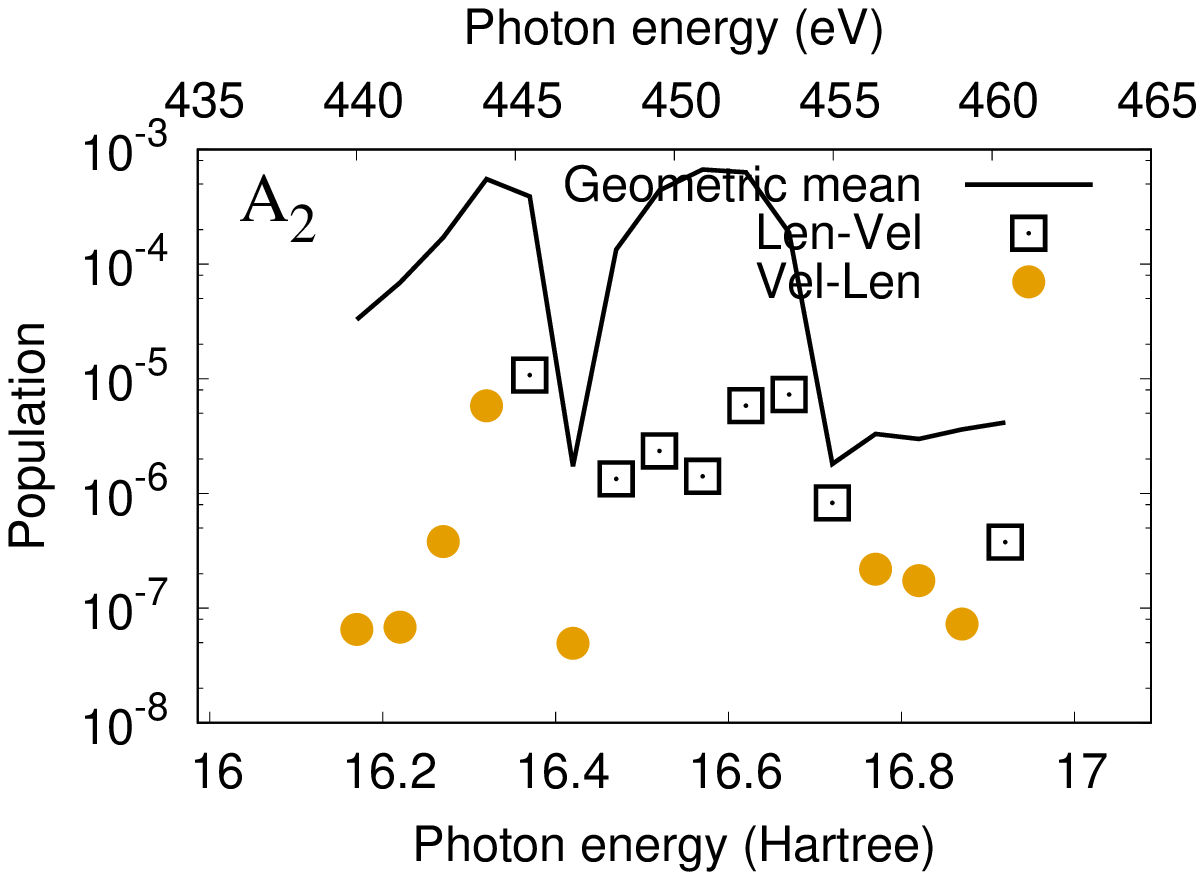}} &
\resizebox{0.66\columnwidth}{!}{\includegraphics*[0.6in,0.6in][5.9in,4.2in]{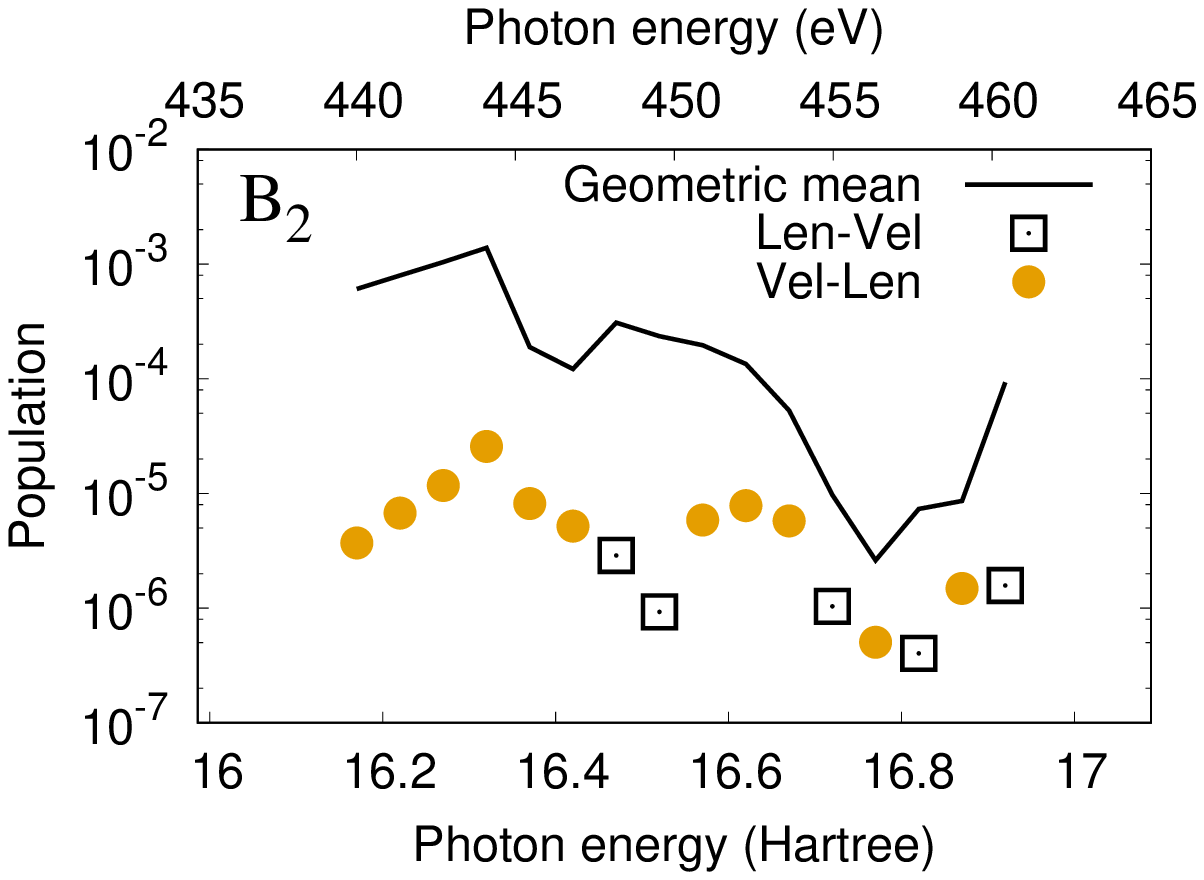}} \\
\end{tabular}
\caption{(Color online) Comparison of population transfer calculated in length and velocity gauge,
at 3.16 $\times$ 10$^{17}$ W cm$^{-2}$, calculated at the larger 0.2975614$a_0$ grid resolution,
presented in order to facilitate a judgment of the effect of discretization error and the convergence of the result with 
respect to the spacing between adjacent sinc basis functions.  The geometric means of the length and velocity gauge
population transfer results are plotted along with their differences.  Black squares and grey (orange) dots indicate
that the sign of the difference between length and
velocity is positive and negative, respectively.
\label{gaugefig}}
\end{figure*}

\begin{figure*}
\begin{tabular}{ccc}
\resizebox{0.66\columnwidth}{!}{\includegraphics*[0.6in,0.6in][5.9in,4.2in]{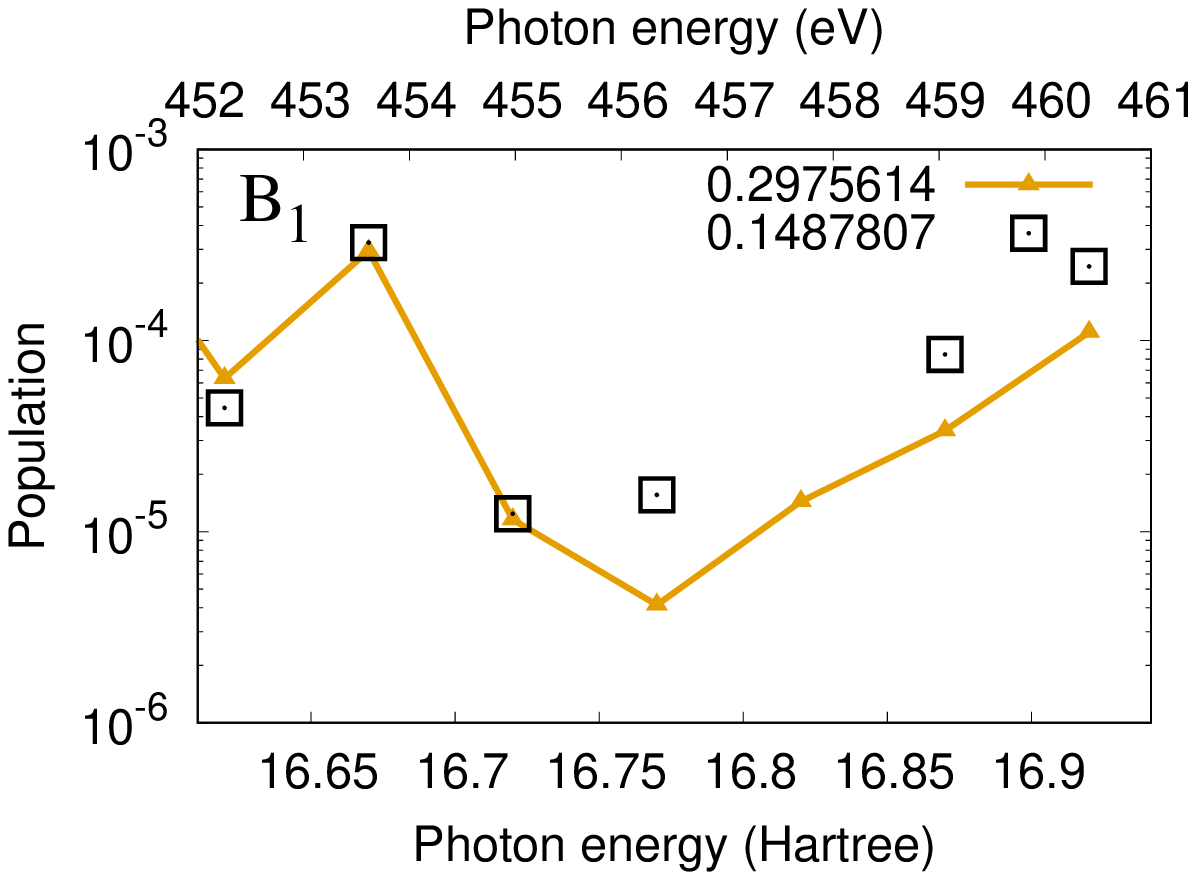}} &
\resizebox{0.66\columnwidth}{!}{\includegraphics*[0.6in,0.6in][5.9in,4.2in]{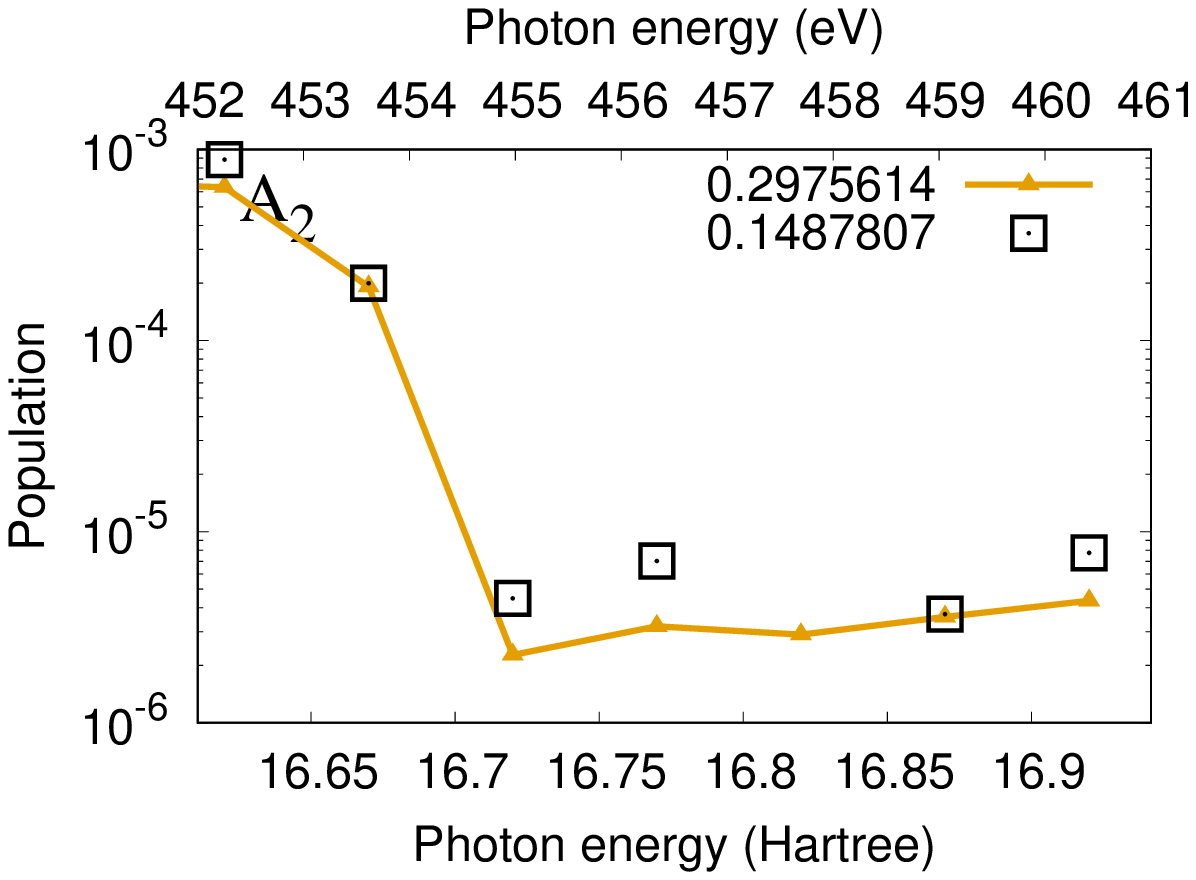}} &
\resizebox{0.66\columnwidth}{!}{\includegraphics*[0.6in,0.6in][5.9in,4.2in]{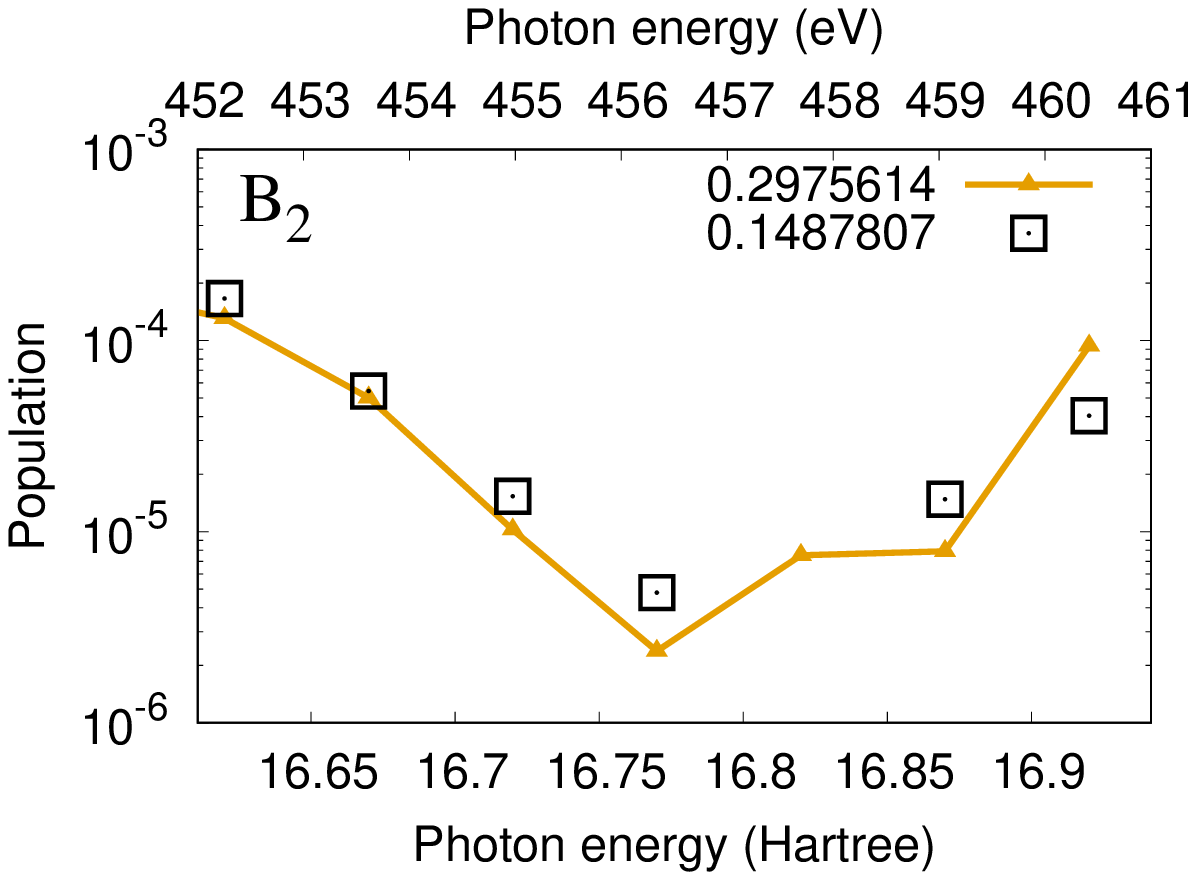}} \\
\end{tabular}
\caption{(Color online) Convergence with respect to grid resolution.  The populations for
the calculation with polarization vector 45 degrees relative to the principal axes of the molecule, i.e. not averaged
over orientations, calculated
at 3.16 $\times$ 10$^{17}$ W cm$^{-2}$, are shown at two resolutions, 0.2975614$a_0$ grid resolution like 
most of the calculations in this paper, and the smaller 0.1487807$a_0$, with grid sizes 55x55x55 and 111x111x111.
\label{111fig}}
\end{figure*}

\begin{figure*}
\begin{tabular}{ccc}
\resizebox{0.66\columnwidth}{!}{\includegraphics*[0.6in,0.6in][5.9in,4.2in]{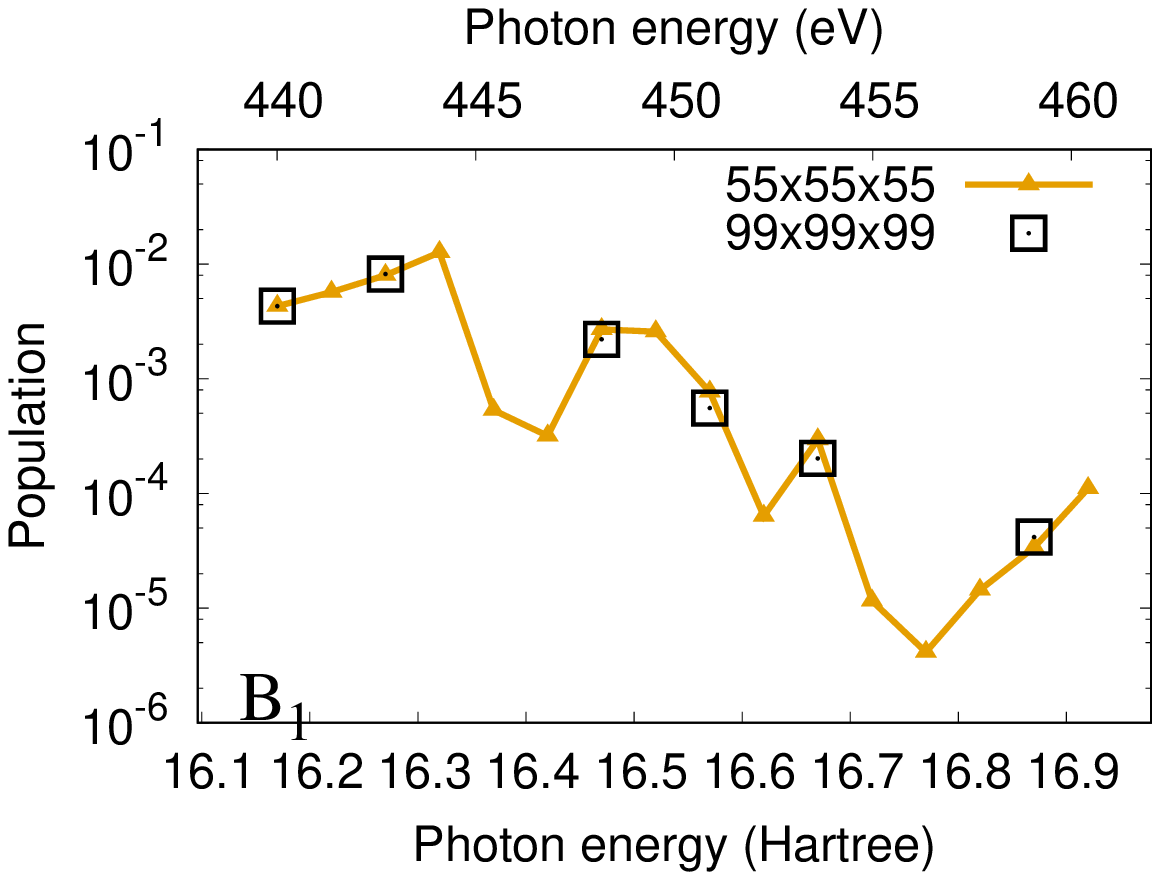}} &
\resizebox{0.66\columnwidth}{!}{\includegraphics*[0.6in,0.6in][5.9in,4.2in]{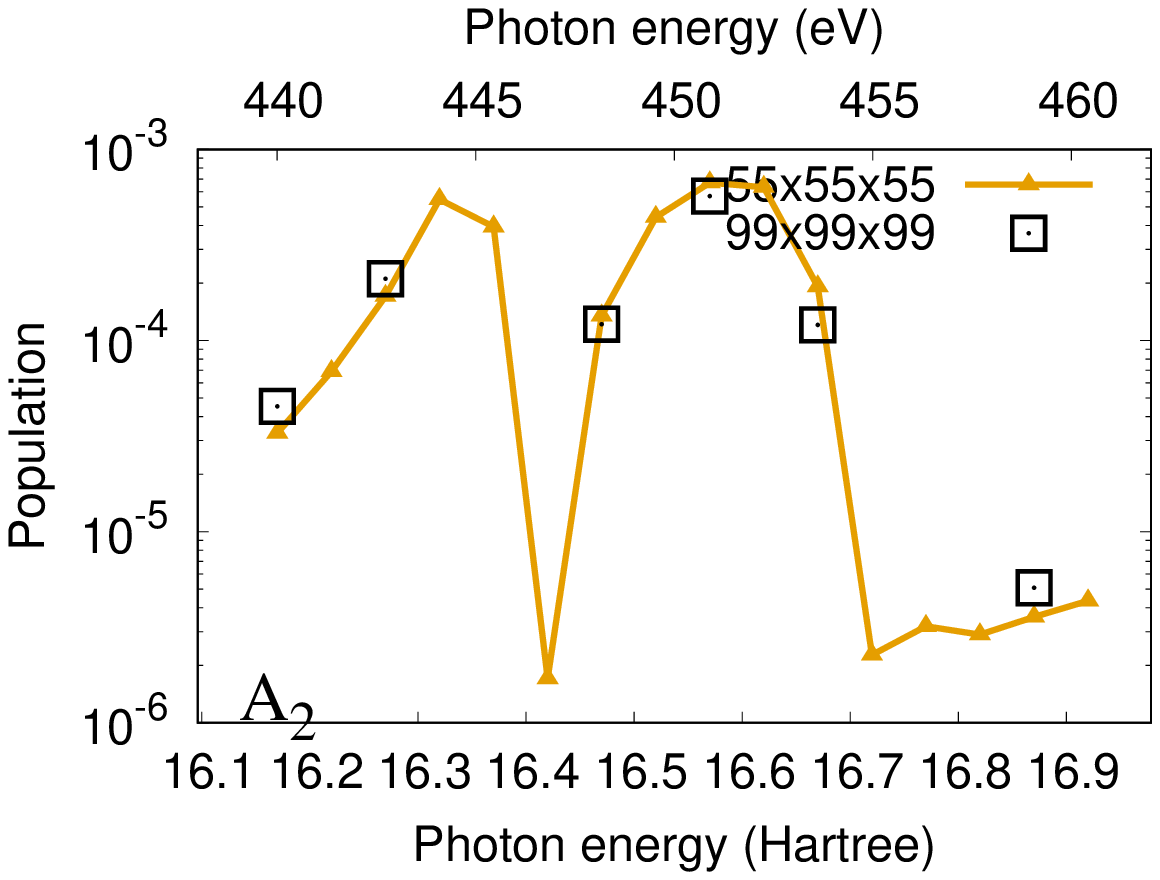}} &
\resizebox{0.66\columnwidth}{!}{\includegraphics*[0.6in,0.6in][5.9in,4.2in]{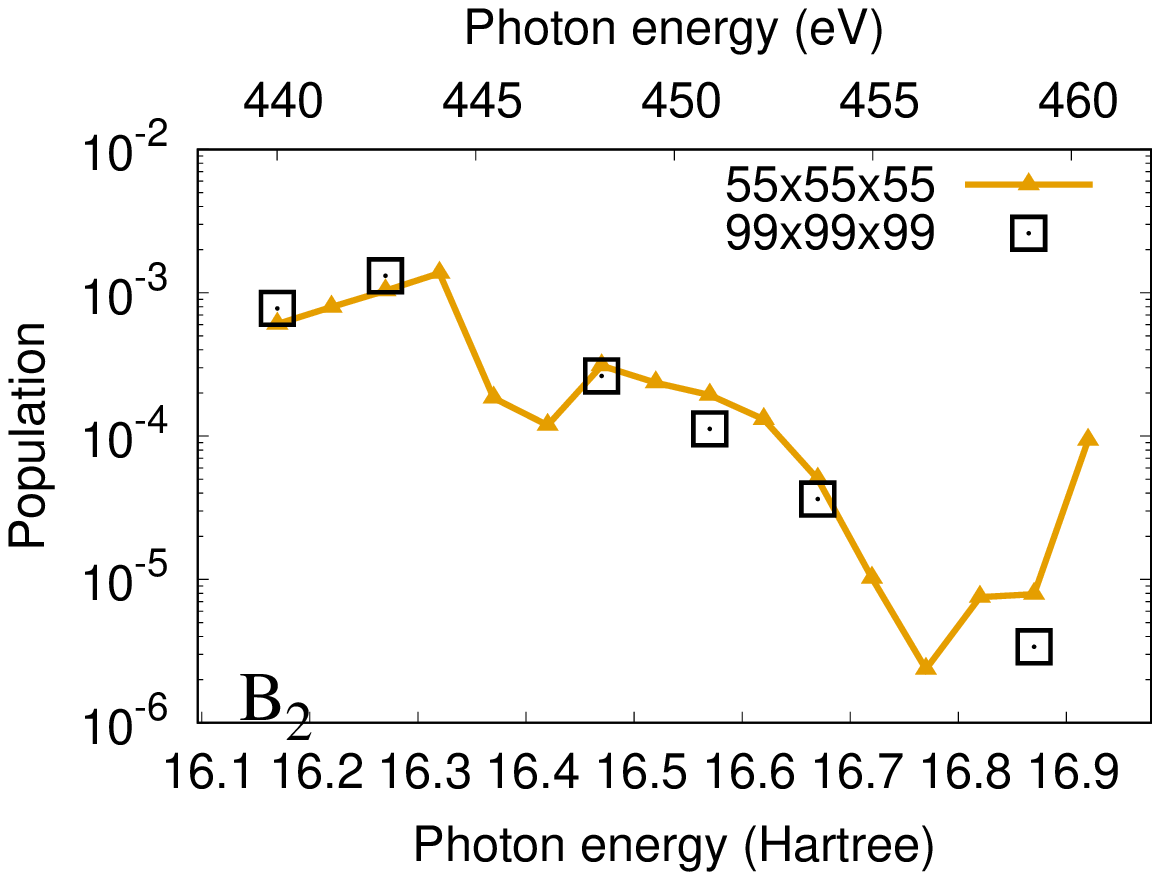}} \\
\end{tabular}
\caption{(Color online) Convergence with respect to primitive sinc DVR basis set, at fixed resolution.
 The populations for
the calculation with polarization vector 45 degrees relative to the principal axes of the molecule, i.e. not averaged
over orientations, calculated
at 3.16 $\times$ 10$^{17}$ W cm$^{-2}$ using the 55x55x55 basis used for most of this paper, 
are compared with populations calculated with a larger 99x99x99 basis and larger complex coordinate
scaling radius, as described in the text.
\label{gridfig}}
\end{figure*}

\section{Photoionization cross section}

Fig.~\ref{xsectfig} shows 
the photoabsorption cross section calculated in the neighborhood of the Oxygen K-edge.
This figure demonstrates that the 
the one-photon transition amplitudes that drive the Raman process are accurately reproduced, although there are 
some significant shifts in the positions of the peaks corresponding to autoionizing states that comprise the near-edge
fine structure.

The bottom panel of this figure shows a higher-resolution result superimposed on the result from the lower-resolution calculation
that is used for the majority of the calculations in this paper.  The resolution is doubled to 0.1487807 bohr, with a 111x111x111 
sinc DVR basis set.  This comparison shows that the excitations are accurately reproduced at the lower resolution, except for the
substantial overall shift in the x-axis values.  The higher-resolution calculation appears to accurately reproduce the location of the 
experimentally-observed edge [CITE], which is obscured and broadened by the overlapping fine structure and by nuclear motion,
around 535eV.  The shift that is required to make the results at the higher and lower resolutions coincide, the shift used in
defining the x-axis ranges in the bottom panel of Fig.~\ref{xsectfig}, is 2.42 Hartree, approximately 66eV.  We use this shift, 2.42 Hartree
or 66eV, when defining shifted numbers throughout the remainder of the text.  We report the energies from the lower-resolution 
calculation, and include parenthetical italicized shifted numbers, using 2.42 or 66eV.  For instance,
``the excitation energy of the XX to YY is ZZZ (\textit{xxxx}) eV.''

The magnitude of the cross section above and below the edge (about 0.5 and a bit more than 1.0 respectively) agree well
with figure 5.10 in Berkowitz's compilation~\cite{berkowitz}.  
The three peaks at about 450 ($A_1$), and 452 \& 453 ($B_1$) correspond to excitations to $6a_1$ and 
$2b_1$ from the Oxygen 1s $\sigma_g$ orbital, and correspond with the peaks observed at approximately 530, 532, and 533 in 
experiment~\cite{Jurgen,Gejo,Piancast}.  Relative to these peaks, there is also a pair of $B_2$ states, at about 462eV, 
both spin couplings for excitation to $5b_2$, and the K-edge lies at about 468-470eV.  It is clear that the K-edge is too high in energy,
and the $B_2$ states are too low, because the $B_2$ states are observed as a broad core-excited shape resonance in experiment.
Experiment~\cite{Jurgen,Gejo,Piancast} gives a $A_1$ to $B_2$ excitation energy of 15eV, and a K-edge about 12eV above  
$A_1$; here they are found at about 12 and 18eV, respectively.  

Because the shift in
the Oxygen K-edge means that the Oxygen and Nitrogen K-edges are unphysically close, one would expect that
the loss due to ionization of the Nitrogen 1$s$ electron would be enhanced and therefore that the population transfer
would be underestimated as a result of this discretization error.  However, the cross sections above
and below the Oxygen K-edge have the correct magnitude, so the effect of the artificial shift in the Oxygen K-edge may
in fact be small.  The higher-resolution $111 \times 111 \times 111$ calculations shown in the bottom panel of Fig.~\ref{xsectfig} demonstrate that all aspects
of the first-order photoionization result are converged at the lower resolution, except for the substantial shift in the K-edge
energy.


\section{Further tests of convergence with respect to single-electron basis \label{convsect}}

\begin{figure*}
\begin{tabular}{ccc}
\resizebox{0.66\columnwidth}{!}{\includegraphics*[0.6in,0.6in][5.9in,4.2in]{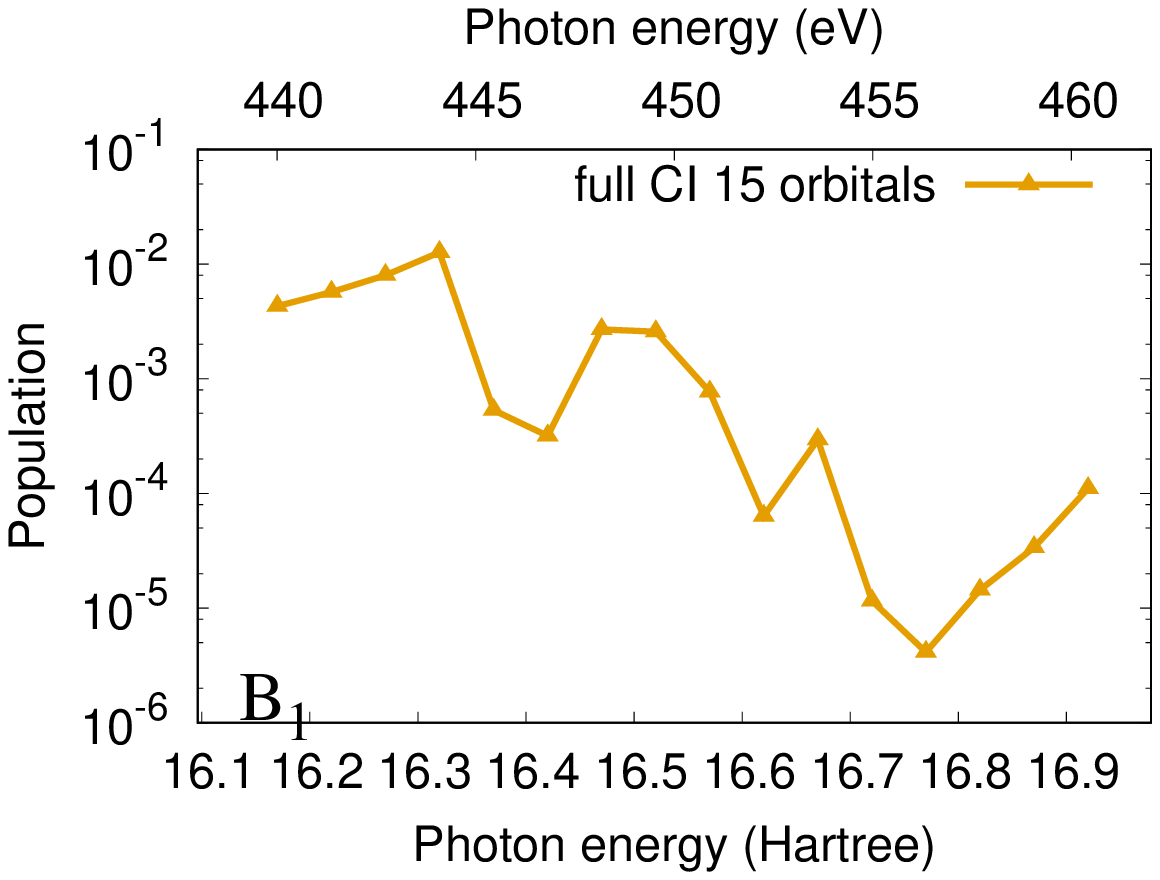}} &
\resizebox{0.66\columnwidth}{!}{\includegraphics*[0.6in,0.6in][5.9in,4.2in]{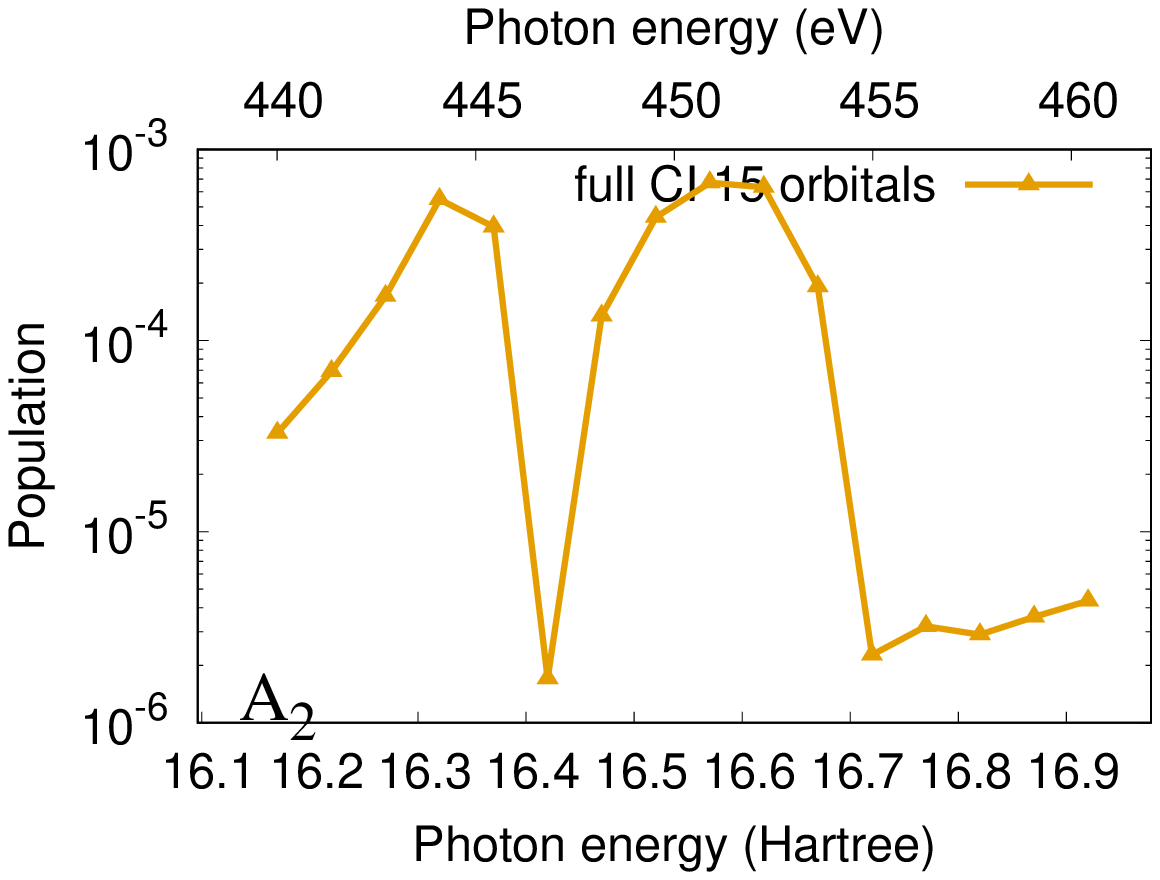}} &
\resizebox{0.66\columnwidth}{!}{\includegraphics*[0.6in,0.6in][5.9in,4.2in]{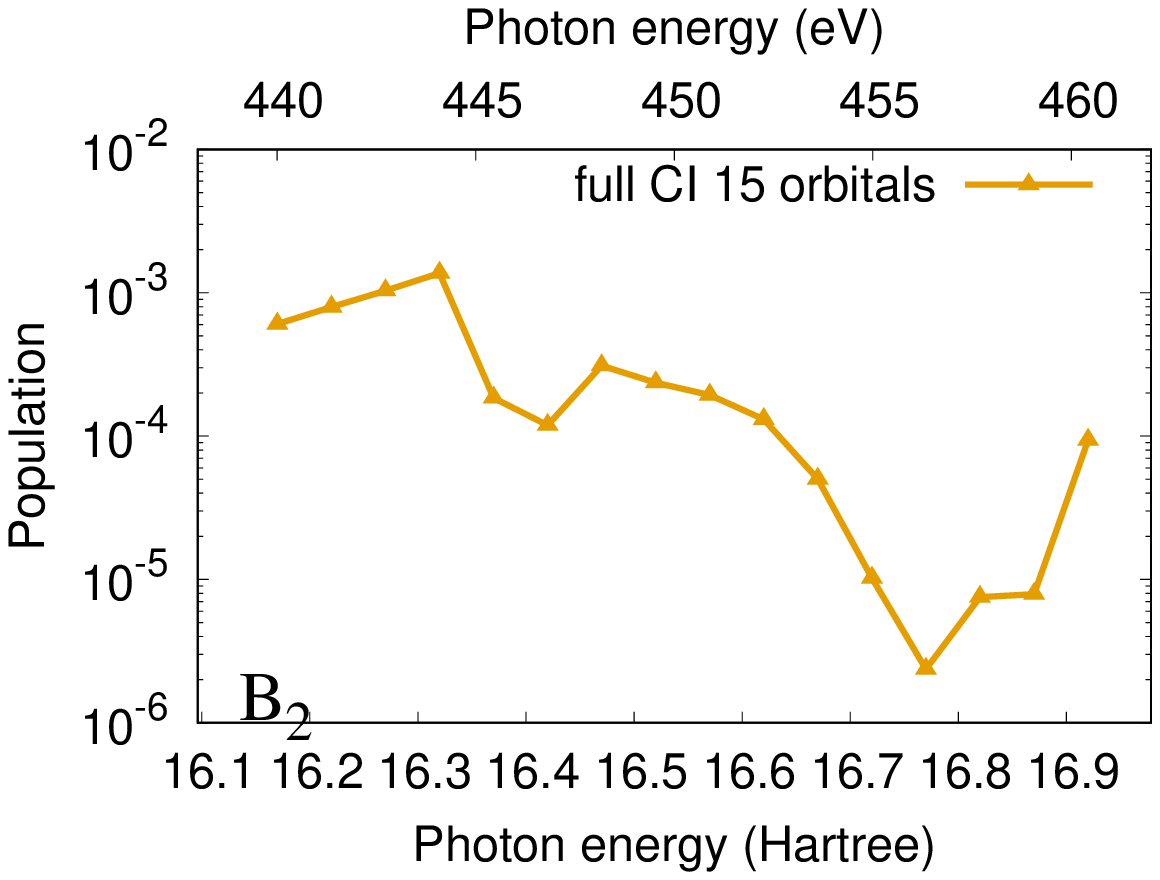}} \\
\end{tabular}
\caption{(Color online)  Convergence with respect to Slater determinant (many-electron) representation.
The populations for
the calculation with polarization vector 45 degrees relative to the principal axes of the molecule, i.e. not averaged
over orientations, calculated
at 3.16 $\times$ 10$^{17}$ W cm$^{-2}$ using the 55x55x55 basis used for most of this paper, 
are compared with populations calculated with the alternate
orbital and Slater determinant basis described in Sec.~\ref{restrictedsect}.
\label{restrictedfig}}
\end{figure*}

The main approximations in these calculations are the omission of 
nuclear motion, the discretization via sinc functions, and the omission
of relativistic (broadly speaking, non-dipole) effects.  The last is the most severe, and will be tested in subsequent work.
Otherwise, the results for population transfer with fixed nuclei should closely 
correspond to the physical result for electronic state population transfer, because the nitrogen and oxygen atoms are 
expected to move very little over the one-femtosecond duration of the pulse.   At the limits of intensity, non-dipole effects may become
significant and future work will examine their effect on impulsive Raman transitions.  

As discussed in the section above, the error due to discretization (grid resolution) appears to be small, based on the 
first-order results, beyond the substantial shift in the Oxygen K-shell 1$s$ ionization energy.  The first-order behavior
is converged, except for the shift.  The spacing of the sinc functions
(0.2975614 $a_0$) is sufficient to represent plane waves with an energy up to 6 keV, so the wave function in the asymptotic
region does not suffer from any discretization error.  The only error due to discretization is due to the cusps in the electronic
wave function at the nuclei.  

In order to more fully judge the effect of the discretization error at the cusps at the nuclei for the nonlinear
population transfer process,
we compare the results of length and 
velocity gauge in Fig~\ref{gaugefig}.  As the spacing between adjacent sinc basis functions
goes to zero, the difference between length and velocity gauge also goes to zero.  Therefore,
the difference between length and velocity gauge gives error bars for the results.  

Velocity gauge 
calculations place a greater emphasis on the wave function near the nuclei, whereas length
gauge calculations place greater weight on the long-range wave function.  Therefore, we regard
length gauge calculations to be more reliable.

However, the issue is moot because, like the lowest-order photoionization results presented in the section above,
the agreement between length and velocity gauge for the nonlinear population transfer
result, demonstrated in Fig~\ref{gaugefig}, is quite good.
Given the large spacing between sinc functions -- again, 0.2975614 $a_0$ -- this agreement may be surprising.  The sinc
basis functions do not accurately represent the cusp at the nucleus, and as demonstrated in Fig.~\ref{xsectfig}, the 
position of the Oxygen K-edge has a substantial error.  However, despite these considerations, the results for population
transfer in Fig.~\ref{gaugefig} in
length and velocity gauge in Fig have an agreement of better than one part in ten (10\%) to 1000 (0.1\%) .

As demonstrated in Ref.~\cite{sincdvr}, the discrete variable representation (DVR) approximation to the Coulomb matrix elements,
involving the inverse of the kinetic energy matrix, gives results that are far superior to the variational method.  Because this DVR
approximation is not variational, considerations based on the smoothing of the cusp at the nucleus do not necessarily apply.
As demonstrated in Ref.~\cite{sincdvr}, 
much better results are obtained with the nonvariational DVR approximation in Ref.~\cite{sincdvr} than with the variational method.
The DVR approximation for the Coulomb matrix elements described in Ref.~\cite{sincdvr} preserves the relationship between the 
kinetic energy and the Coulomb operator -- the Coulomb operator is the integral kernel of the inverse of the kinetic energy
operator -- in matrix form.  We speculated that the preservation of this relationship is responsible for the surprising accuracy
of virial theorem ratios presented in Ref.~\cite{sincdvr}, and we also speculate that it is responsible 
 for the surprising agreement between length and velocity gauge presented in Fig.~\ref{gaugefig}.

We present an explicit demonstration of convergence with respect to grid resolution in Fig.~\ref{111fig}.  The resolution
is doubled and everything else is kept constant.  
The spacing is halved to 0.1487807$a_0$ and the grid is doubled to $111 \times 111 \times 111$.
Only one calculation for the orientation average is shown, the one corresponding
to the Lebedev quadrature point for which the polarization vector is parallel with the vector $(x,y,z)=(1,1,1)$, in which $x, y, z$ are
principal (c$_{2V}$) axes of the molecule, and only a few calculations have been performed due to limited computer resources.
These few calculations however show good convergence, well within an order of magnitude.  
Better convergence might be found by slightly adjusting the intensity, because small differences in first-order
transition strengths are magnified in a nonlinear process.
Further work is required for a more complete study of the 
convergence with respect to grid resolution, including the orientation average required to predict the observed result.

In Fig.~\ref{gridfig}, we demonstrate the convergence with respect to the sinc DVR orbital basis, keeping the resolution 
fixed.  Again, only the polarization vector $(x,y,z)=(1,1,1)$ calculation is performed, not the entire orientation average.
The convergence with respect to the extent of the grid and the exterior smooth complex coordinate scaling is tested by 
comparing the results from the $55 \times 55 \times 55$ calcuation
with complex scaling starting at $x$, $y$, or $z=4a_0$, with results obtained from a larger
$99 \times 99 \times 99$ calculation with complex scaling starting at $x=a_0$ and extended to $x=14a_0$,
in which $x$ is the real-valued coordinate that parameterizes the complex-valued ray.  
The stretching factor and scaling angle were kept
at their previous values of 3 and 0.5 radians (about 60 degrees).  The agreement of the results is good, with ten percent.

\begin{figure*}
\begin{tabular}{ccc}
\resizebox{0.66\columnwidth}{!}{\includegraphics*[0.6in,0.6in][5.9in,4.2in]{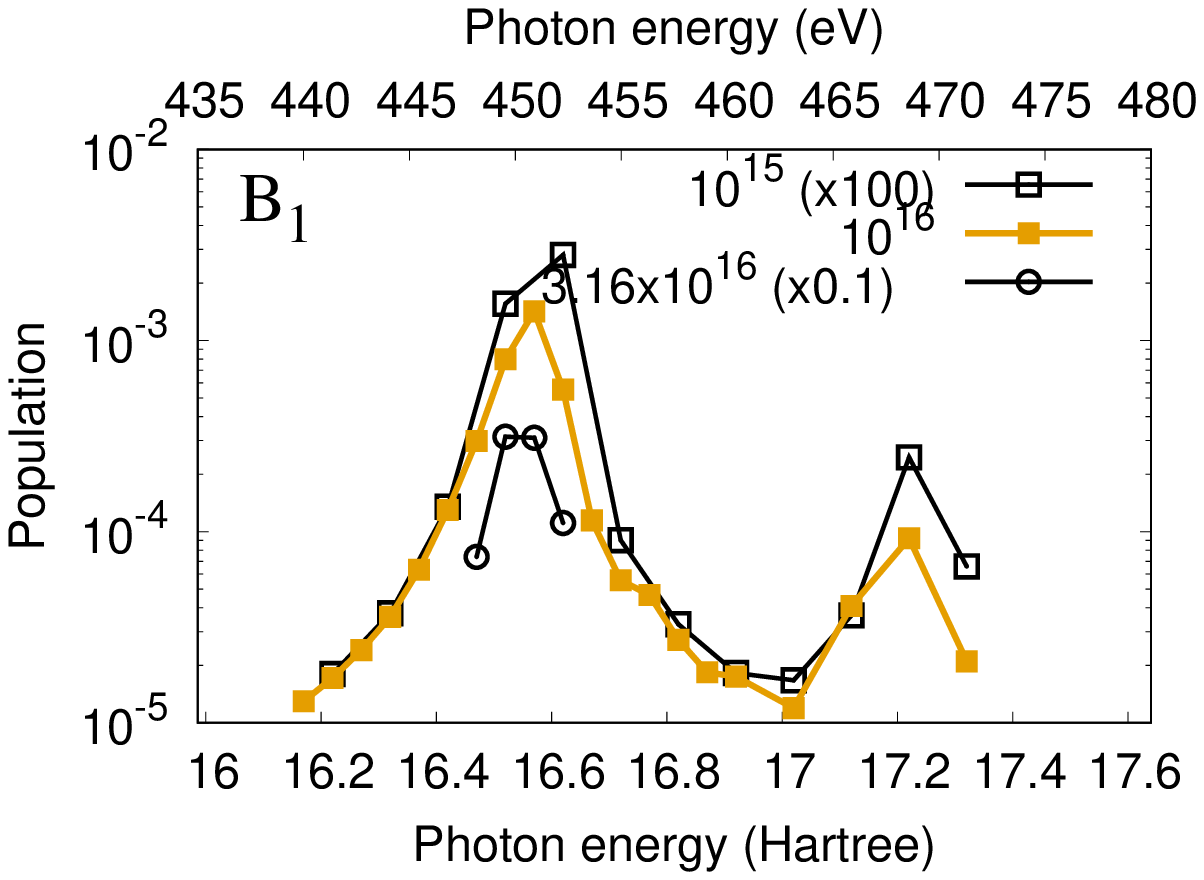}} &
\resizebox{0.66\columnwidth}{!}{\includegraphics*[0.6in,0.6in][5.9in,4.2in]{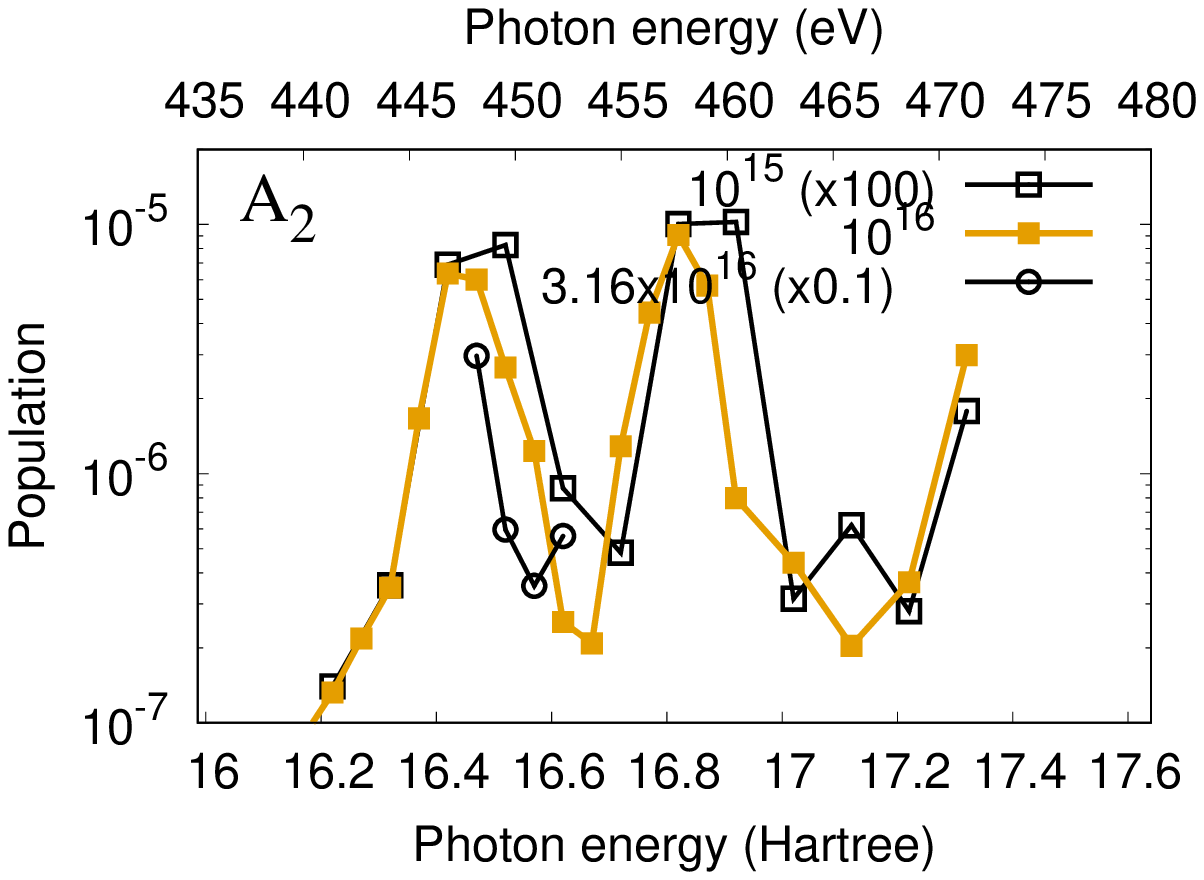}} &
\resizebox{0.66\columnwidth}{!}{\includegraphics*[0.6in,0.6in][5.9in,4.2in]{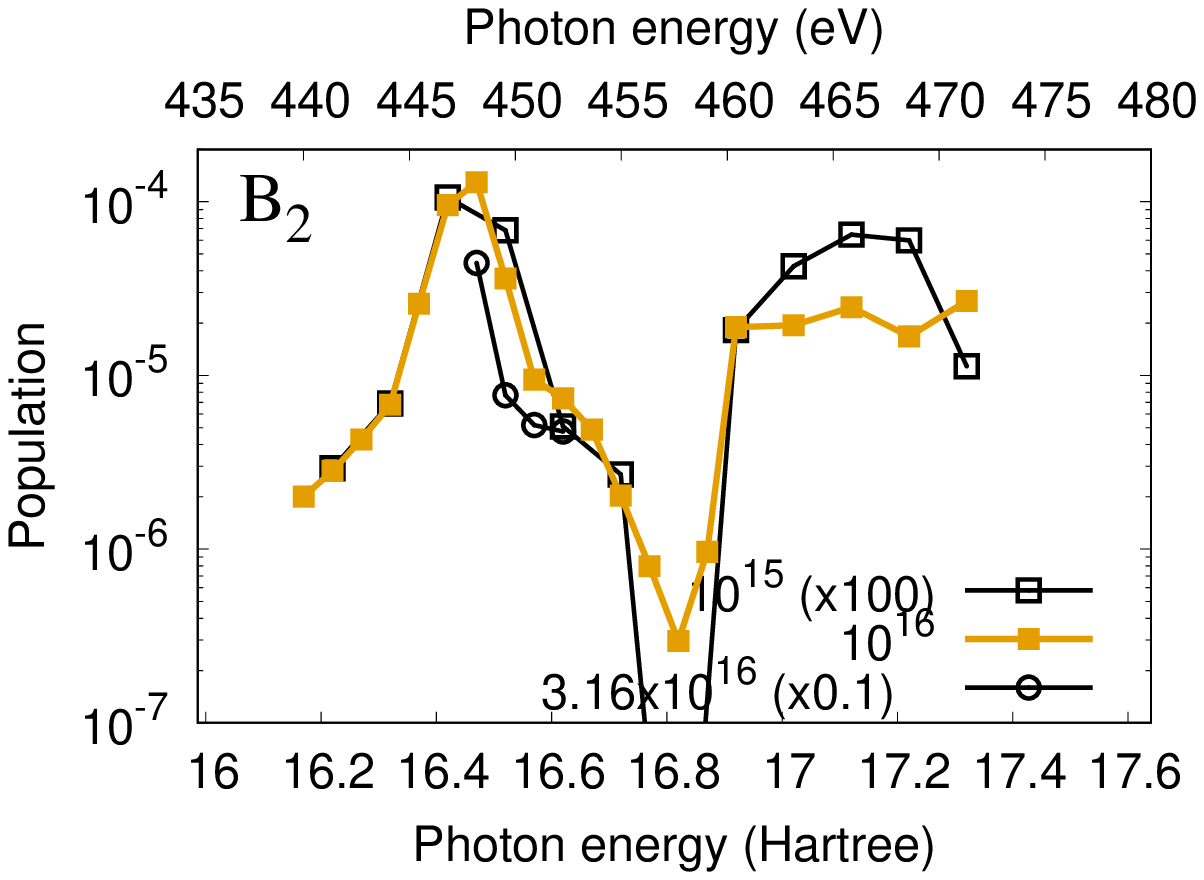}} \\
\end{tabular}
\caption{(Color online) Behavior of population transfer at low intensity.  
Different intensities are plotted with different lines and labeled
in Watts per square centimeter.
The results at 10$^{15}$ and 3.6 $\times$ 10$^{16}$ Watts per square centimeter
have been multiplied by factors to facilitate a comparison of the generalized two-photon cross section.  To obtain
a generalized two-photon cross section in megabarns squared $\times$ femtoseconds, multiply the plotted y-axis value by 345 times
the x-axis value squared.
\label{lowfig}}
\end{figure*}

\section{Convergence with respect to many-electron basis \label{restrictedsect}}

The many-electron convergence is tested by varying the orbital and Slater determinant basis.
The MCTDHF method employs time-dependent orbitals and therefore
every electron can be simultaneously excited, regardless of the Slater determinant space.
The defects imposed upon the time-dependent MCTDHF 
solution by the truncation of the Slater determinant space are not yet understood, and
different in nature from those due to truncated time-independent Slater determinant basis sets.

Furthermore, the nature of these defects must be intimately tied to the chosen variational principle.
The coupling terms among orbitals are dictated by the chosen variational principle, and these coupling terms
distinguish different MCTDHF methods involving truncated Slater determinant spaces.  However, these terms are
often approximated [CITE,CITE] are entirely neglected [CITE,CITE].  We have provided a derivation of explicit working
equations for the McLachlan and Lagrangian variational principles~\cite{restricted}.  Those
working equations are employed in the tests reported here -- specifically, the method that mixes
10\% McLachlan with 90\% Lagrangian.  

In general, we find~\cite{restricted} that the time-independent many-electron basis sets used in 
contemporary MCSCF (multiconfiguration self-consistent field) and configuration-interaction,
and also in configuration-interaction representations of time-dependent problems, do not behave
well in the MCTDHF method~\cite{restricted}, following either the Lagrangian (least action) or McLachlan (minimum norm
of the error) stationary principle.  Symmetry breaking is more severe~\cite{restricted} 
using typical Slater determinant
sets such as CISD (configuration interaction with single and double excitations).  We have found~\cite{restricted}
that the particle-hole conjugates of these basis sets, corresponding to a full configuration outer shell with a few
excitations allowed from the core, perform best.

However, here we expand the orbital basis greatly, from 15 to 20 orbitals, in order to fully test the convergence
with respect to orbitals, and employ a CISD slater determinant space.  We divide the 20 orbitals into two shells of 11 and 9.
The wave function is represented using CIS, single transitions only from the first 11 to the last 9 time-dependent orbitals.
Matrix elements are computed among the CISD determinant space.
18 to 20 electrons are allowed in the first
shell and one to three electrons in the second.
In total, only 1296 spin-adapted linear combinations of Slater determinants
are explicitly included in the CIS representation, and matrix elements among 62940 CISD 
Slater determinants are recomputed each time step.  Essentially all of the time in this calculation is spent
propagating orbitals.

In comparison, for the rest of the calculations in this paper we use full configuration interaction with 23 orbitals,
yielding 621075 spin-adapted linear combinations of Slater determinants.
The two calculations, 15-orbital full configuration and 20-orbital CISD, are very different and agreement
between these calculations would strongly imply convergence with respect to the Slater determinant representation.

In Fig.~\ref{restrictedfig}, we show

BLAH

BLAH

BLAH

BLAH

BLAH

BLAH

BLAH

\section{Low-intensity behavior}


The behavior of the population transfer at relatively low intensity is shown in Fig.~\ref{lowfig}.  This figure shows the population
transfer at intensities for which the second-order behavior begins to break down.  Intensities of
1$\times$10$^{15}$ W cm$^{-2}$, 1$\times$10$^{16}$ W cm$^{-2}$, and 3.16$\times$10$^{16}$ W cm$^{-2}$
are plotted, and the first and last of these are multiplied by factors of 100 and 0.1, respectively, in order to judge the
degree to which second-order behavior is obeyed.  If second-order perturbation theory were to hold for these
intensities, the lines in Fig.~\ref{lowfig} would coincide. 

In this figure one can see that the behavior is still second-order at an intensity of 1$\times$10$^{16}$ W cm$^{-2}$ 
for the lower photon energies, below about 447eV, because the lines corresponding to 1$\times$10$^{16}$ W cm$^{-2}$ and
1$\times$10$^{15}$ W cm$^{-2}$ coincide in that region, on the left side of the figures.  Above 447eV, these two lines
diverge, so it is clear that the behavior at 1$\times$10$^{16}$ W cm$^{-2}$ is beyond second-order when the central
frequency is close and above the near-edge fine structure and Oxygen K-edge.

The low-intensity behavior in Fig.~\ref{lowfig} mirrors what is expected from second-order perturbation theory.  The
optimum population transfer at low energy occurs at about 450eV for the B$_1$ state, and about 447eV for the B$_2$
state.  For the A$_2$ state, for which the population transfer is smaller, there are a pair of local maxima at about 447 and
458eV.  

The mechanisms for population transfer can be inferred by reference to Fig.~\ref{xsectfig}, which shows the photoionization
cross section.  In that figure, there is an A$_1$ metastable state at about 450eV, a pair of B$_1$ states at about 452 and 453eV;
and a pair of B$_2$ states at about 461 and 462eV.  The Oxygen K-edge is obscured by additional peaks corresponding
to metastable states, but the edge seems to occur at about 470eV.

Because the optimum population transfer at low intensity occurs far below the edge, relative to the spectral bandwidth of 3.25eV,
it is clear that the transitions are driven through the metastable, discrete autoionizing states comprising the near-edge fine 
structure at low intensity.  This low-intensity behavior for 1fs FWHM pulses
is consistent with the picture presented in Ref.~\cite{tiger}.

The Hartree-Fock configuration of the initial ground state is 
$1a_1^2 2a_1^2 1b_2^2 3a_1^2 2b_2^2 4a_1^2 3b_2^2 5a_1^2 1b_1^2 4b_2^2 1a_2^2 6a_1^1$.  
The metastable states comprising the near-edge fine structure that are visible as the most prominent peaks 
in Fig.~\ref{xsectfig}, A$_1$, B$_1$, and B$_2$, and which have been observed in experiment~\cite{Jurgen,Gejo,Piancast}
are described as excitations $1a_1\rightarrow 6a_1$, 
$1a_1\rightarrow 2b_1$, and $1a_1\rightarrow 5b_2$.
The valence B$_1$, A$_2$, and B$_2$ states are described as excitations $6a_1 \rightarrow 2b_1$, 
$1a_2 \rightarrow 6a_1$, and $4b_2 \rightarrow 6a_1$.

The optimum at 450eV for population transfer to the valence B$_1$ state in Fig.~\ref{lowfig} is consistent with
two-photon transitions via the intermediate B$_1$ states, proceeding as $1a_1 \rightarrow 2b_1$, $6a_1 \rightarrow 1a_1$.
In the photoionization cross section in Fig.~\ref{xsectfig}, the two ($1a_1 \rightarrow 2b_1$) $^2B_1$ states at 452 and 453eV
have the same orbital occupancy but different spin parentage.  They
are described most closely as states in which the singly-occupied $2b_1$ and $6a_1$ orbitals are singlet or triplet
coupled.  The triplet-coupled $1a_1^{-1}6a_1^1 2b_1^1$ B$_1$ intermediate state at 453eV, with a smaller oscillator
strength (smaller peak) in Fig.~\ref{xsectfig}, is that which drives the Raman transition.

The A$_1$ state at 450eV ($1a_1 \rightarrow 6a_1$) is not expected to contribute because the subsequent transition
downward would involve the transition of more than one electron in the Hartree-Fock picture.
In Fig.~\ref{xsectfig}
it is clear that the oscillator strength from the ground state to the B$_1$ state at 452eV is greatest among these three,
and the 450eV optimum is most closely consistent with a transition via a 452eV intermediate state, given the 3.25eV
bandwidth.  

The 447eV optimum for population transfer to the valence B$_2$ state in Fig.~\ref{lowfig} is consistent with a transition through the lowest
A$_1$ metastable state at 450eV, proceeding as $1a_1 \rightarrow 6a_1$, $4b_2 \rightarrow 1a_1$.  
A two-photon transition via the intermediate B$_1$ state is not dipole-allowed.  
There is no second maximum corresponding to population transfer via the intermediate B$_2$ state ($1a_1 \rightarrow 5b_2$) 
at higher energy because the Stokes transition would require moving more than
one electron in the Hartree-Fock picture.

The A$_2$ valence state population transfer at low intensity seen in Fig.~\ref{lowfig}
has two maxima, at about 447 and 458eV.  The higher-photon-energy 458eV maximum is slightly greater.  Both of these
maxima are lower for this A$_2$ state (population transfer equal or less than 10$^{-5}$ at 10$^{16}$ W cm$^{-2}$)
than the maxima for B$_1$ (over 10$^{-3}$ at 10$^{16}$ W cm$^{-2}$) and B$_2$ (about 10$^{-4}$ at 10$^{16}$ W cm$^{-2}$).
The lower population transfer for the valence A$_2$ state is consistent with a transition driven by
configuration mixing in the intermediate or final state.  The $1a_2 \rightarrow 6a_1$ configuration of the valence A$_2$ state
cannot be obtained by two subsequent dipole-allowed one-electron transitions involving the 1$s$ electrons. 
Transitions through the lower B$_1$ states at 450 and 452eV ($1a_1 \rightarrow 2b_1$)
and through the B$_2$ states at 461 and 462eV ($1a_1 \rightarrow 5b_2$) are clearly responsible for the population transfer
to this A$_2$ valence state, given the location of the two sharp maxima for population transfer at about 447 and 458eV.
The low value of the population transfer at these maxima is explained by the fact that 
the subsequent transition to the
final $1a_2 \rightarrow 6a_1$ configuration is driven by configuration mixing in these states, the intermediate B$_2$ and B$_1$
or the final A$_2$.  Correlating excitations $1a_2 5b_2 \rightarrow 6a_1 2b_1$ and $1a_2 2b_1 \rightarrow 6a_1 5b_2$ in the
B$_2$ and B$_1$ intermediate state are the most probable explanations of the calculated population transfer.

\begin{figure*}
\begin{tabular}{cc}
\resizebox{1.0\columnwidth}{!}{\includegraphics*[0.6in,0.6in][5.9in,4.2in]{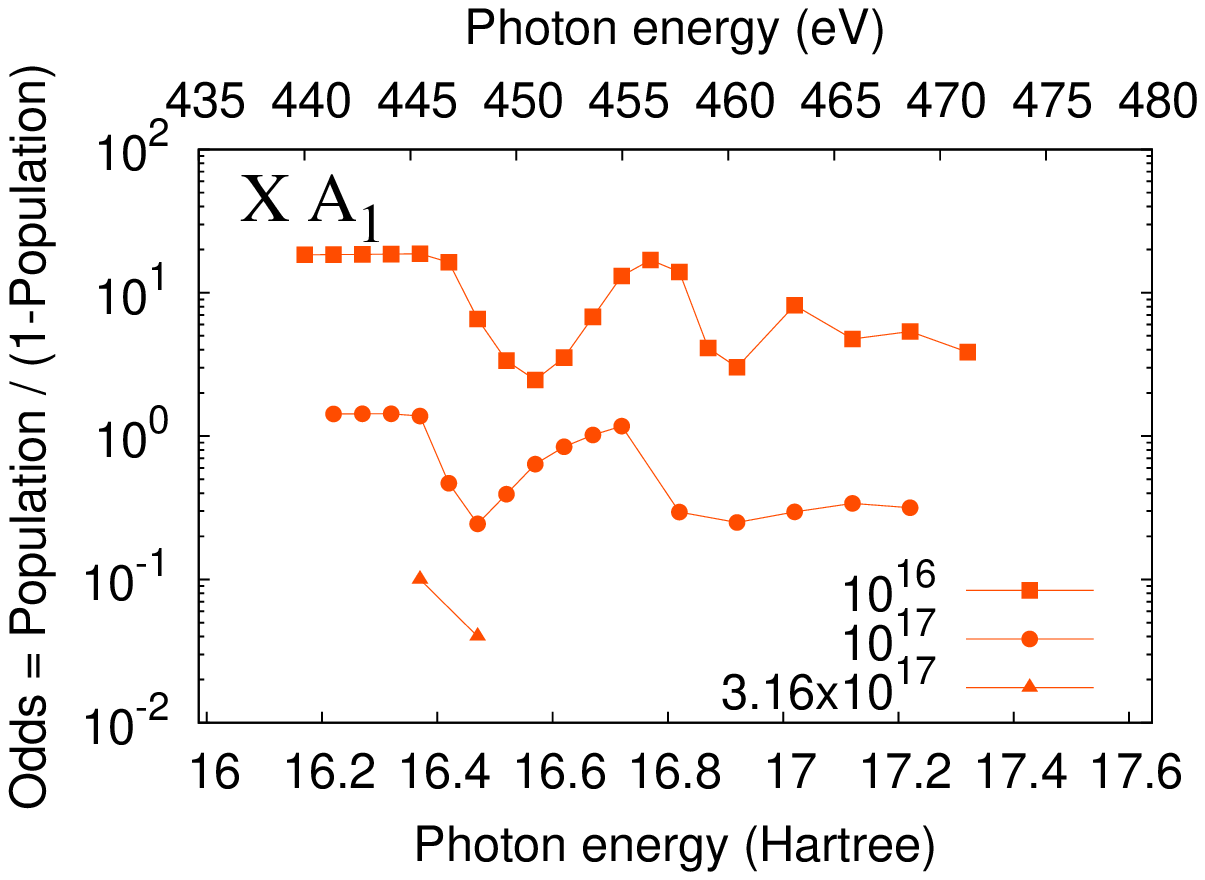}} &
\resizebox{1.0\columnwidth}{!}{\includegraphics*[0.6in,0.6in][5.9in,4.2in]{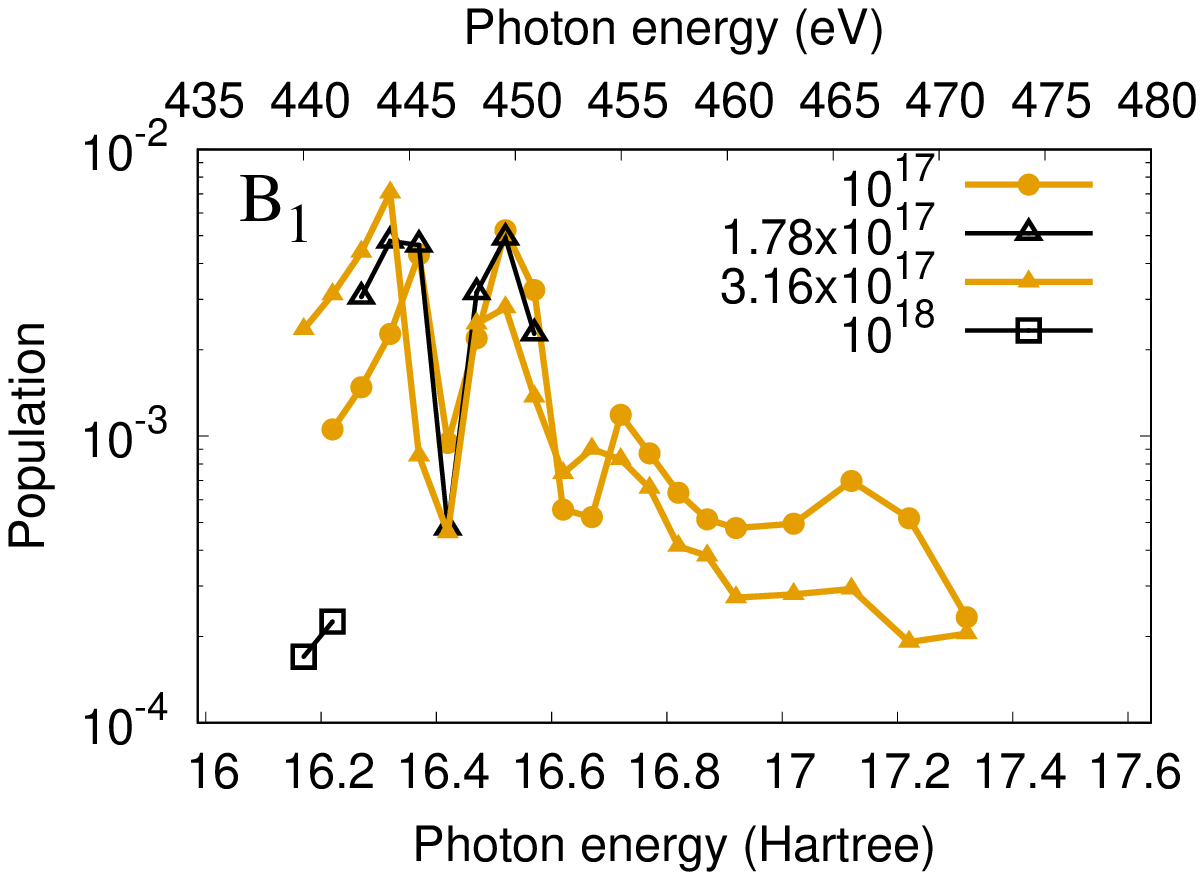}} \\
\resizebox{1.0\columnwidth}{!}{\includegraphics*[0.6in,0.6in][5.9in,4.2in]{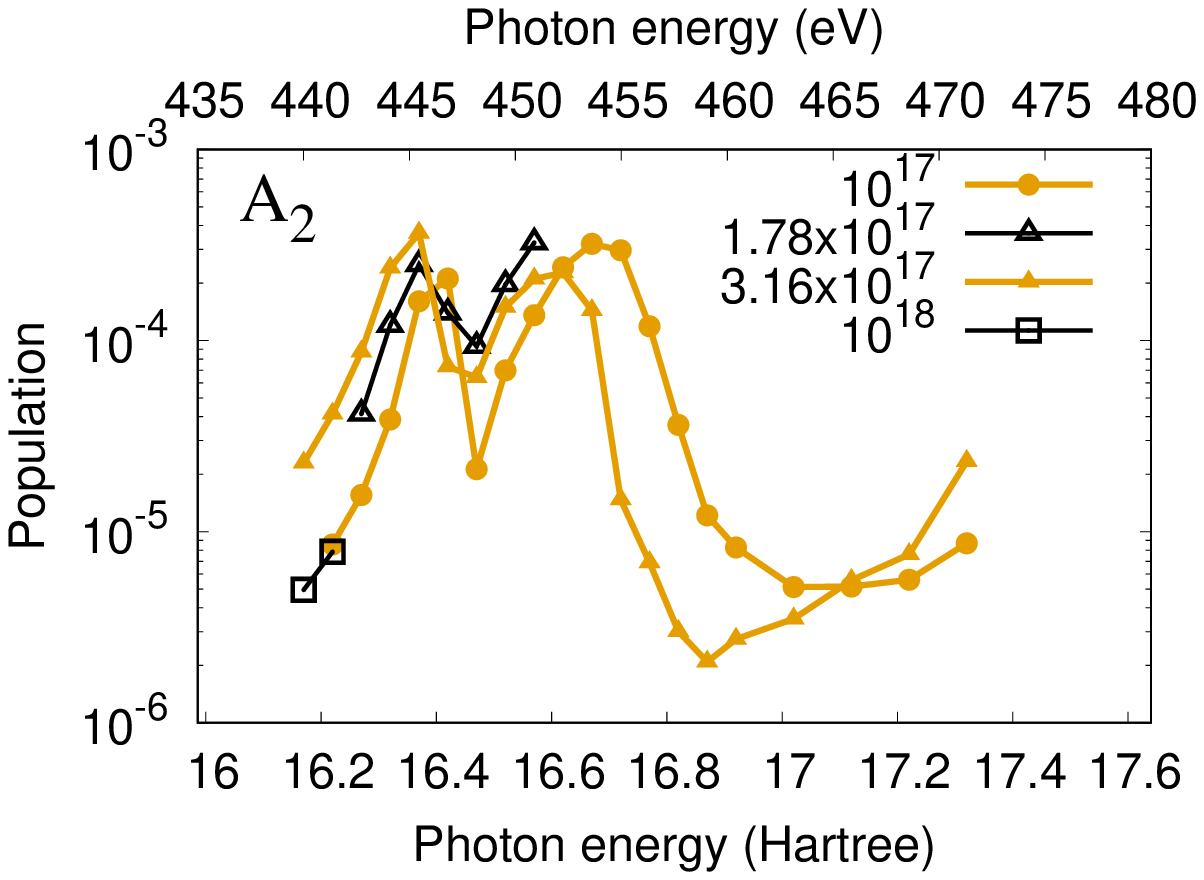}} &
\resizebox{1.0\columnwidth}{!}{\includegraphics*[0.6in,0.6in][5.9in,4.2in]{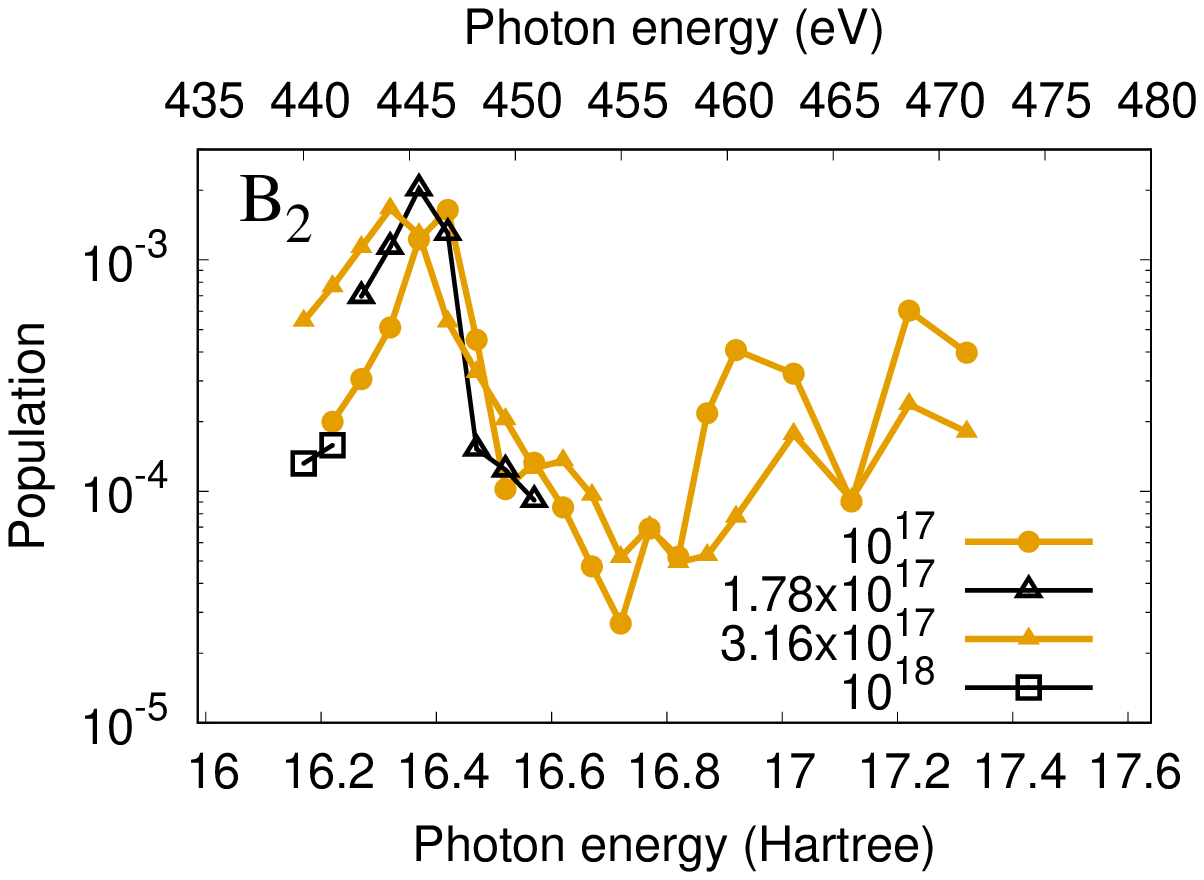}} \\
\end{tabular}
\caption{(Color online) Results for population transfer to the B$_1$, A$_2$, and B$_2$ 
valence excited states of NO$_2$, 
and odds for remaining in the ground state,
averaged over orientation, for 1fs pulses, at higher intensities showing the global optimum population transfer.
Different intensities are plotted with different lines and labeled
in Watts per square centimeter.
\label{popfig}}
\end{figure*}

\section{Optimizing impulsive X-ray Raman excitation of NO$_2$}

These results are limited to linearly polarized 1fs FWHM pulses.  Such pulses are expected to be available
from free-electron lasers in the near future.  In order to further examine the parameter space to find the truly global
optimum for impulsive x-ray Raman population transfer,
we will vary chirp and duration in future work.

The results for population transfer at high intensity, including the global optimum, for these 1fs FWHM linearly polarized
pulses near the Oxygen $1s$ K-edge, are shown in Figure~\ref{popfig}.

%

The best population transfer is obtained for the B$_1$ state, 
at the intensity 3.16$\times$10$^{17}$ W cm$^{-2}$ and approximately 444.1eV (16.32 Hartree) photon energy, substantially
6eV red-detuned from the second-order optimum which occurred at 450eV as described above.  
However the optimum population transfer is only 0.70\% with these one-femtosecond,
linearly polarized pulses, for the orientation average. Among the seven distinct Lebedev quadrature points required to compute the 50th order
orientation average, the largest population transfer to the B$_1$ state was at this intensity and central frequency, 
3.16$\times$10$^{17}$ W cm$^{-2}$ and 6.32 Hartree, at 2.39\%.  This optimum was obtained for the fixed orientation
with a polarization vector
$(x,y,z)=(0,1,1)$ in the molecular frame, in which $(x,y,z)$ are the principal axes of the molecule and $x$ is perpendicular to the plane
of the molecule.

In a well-designed
multidimensional spectroscopy experiment, the efficiency of subsequent steps benefits from the orientation
imparted by prior steps.  The 2.39\% population transfer maximum, greatest among the orientations that we calculated,
 implies a coherence as great as 15\%, which
would easily enable proposed multidimensional attosecond x-ray spectroscopy experiments.  The cleanest experiment
would be performed at a lesser intensity, to reduce noise.  These calculations indicate that there is no fundamental limitation
to the implementation of proposed multidimensional attosecond x-ray spectroscopy methods, based on the efficiency of
the population transfer with 1fs pulses.

Examining the population transfer for B$_1$, going from low intensity in Fig.~\ref{lowfig} to high intensity in Fig.~\ref{popfig},
one can see that a minimum develops at about 446eV.  The peak population transfer does not shift monotonically downward 
as intensity is increased.
Instead, this sharp minimum at about 446eV develops, the original peak at 450eV saturates at about 1.78$\times$10$^{17}$ W cm$^{-2}$, 
and the peak at lower photon energies around 444eV increases to become the global maximum at 3.16$\times$ 10$^{17}$ W cm$^{-2}$.

Global optimum population transfer to the B$_2$ valence state occurs at slightly lower intensity, 1.78$\times$10$^{17}$ W cm$^{-2}$,
at about 445eV, with a population transfer of about 0.2\%,
whereas the second-order optimum was at 447eV.  Going from low to high intensity, the peak population transfer
for the B$_2$ state shifts monotonically downward in energy.

\begin{figure*}
\begin{tabular}{ccc}
\resizebox{0.66\columnwidth}{!}{\includegraphics*[0.6in,0.6in][5.9in,4.2in]{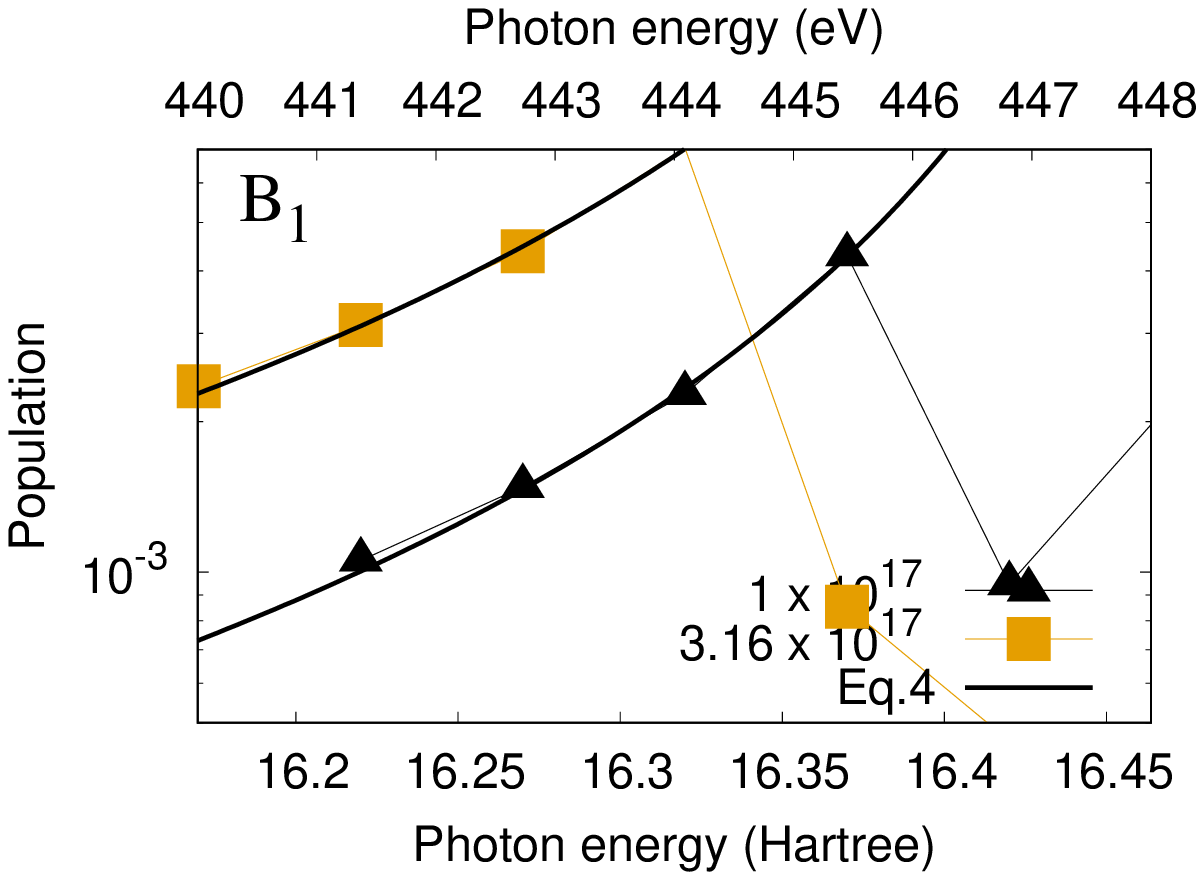}} &
\resizebox{0.66\columnwidth}{!}{\includegraphics*[0.6in,0.6in][5.9in,4.2in]{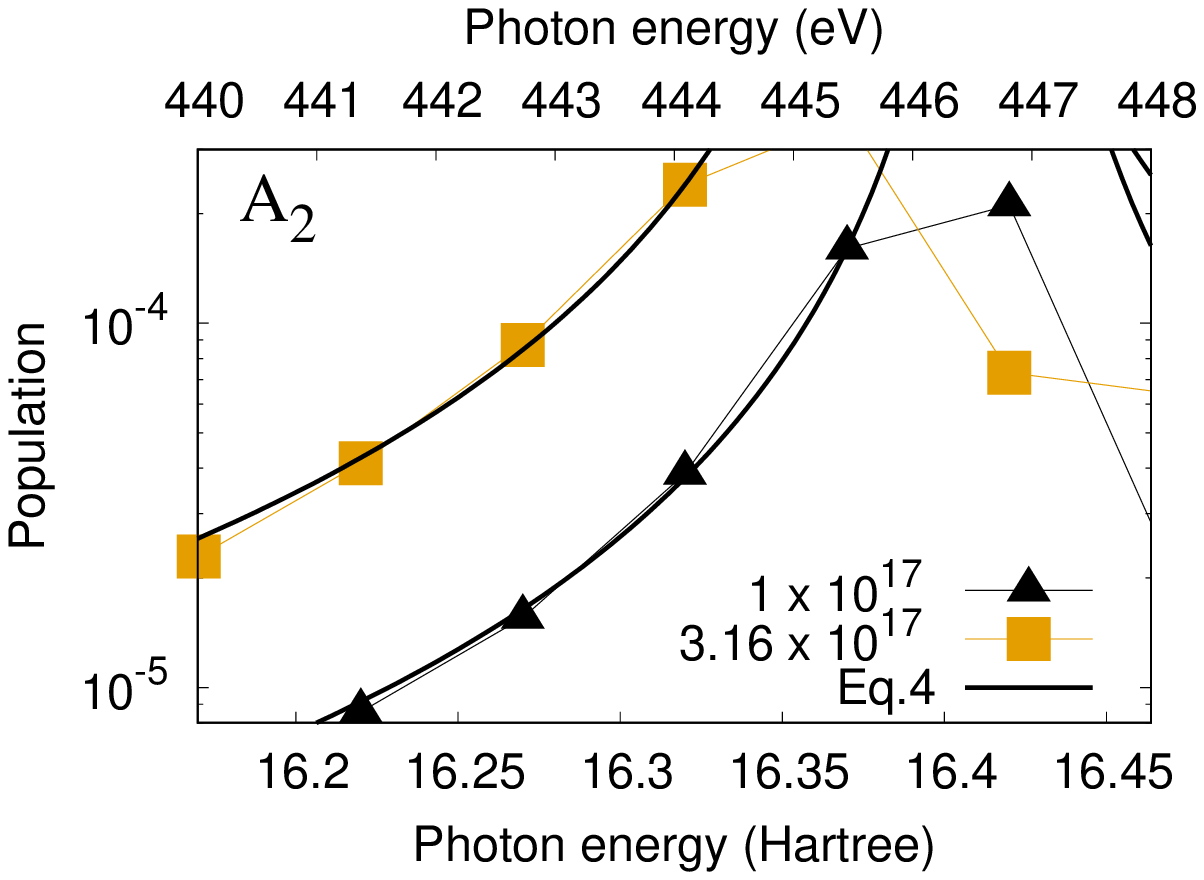}} &
\resizebox{0.66\columnwidth}{!}{\includegraphics*[0.6in,0.6in][5.9in,4.2in]{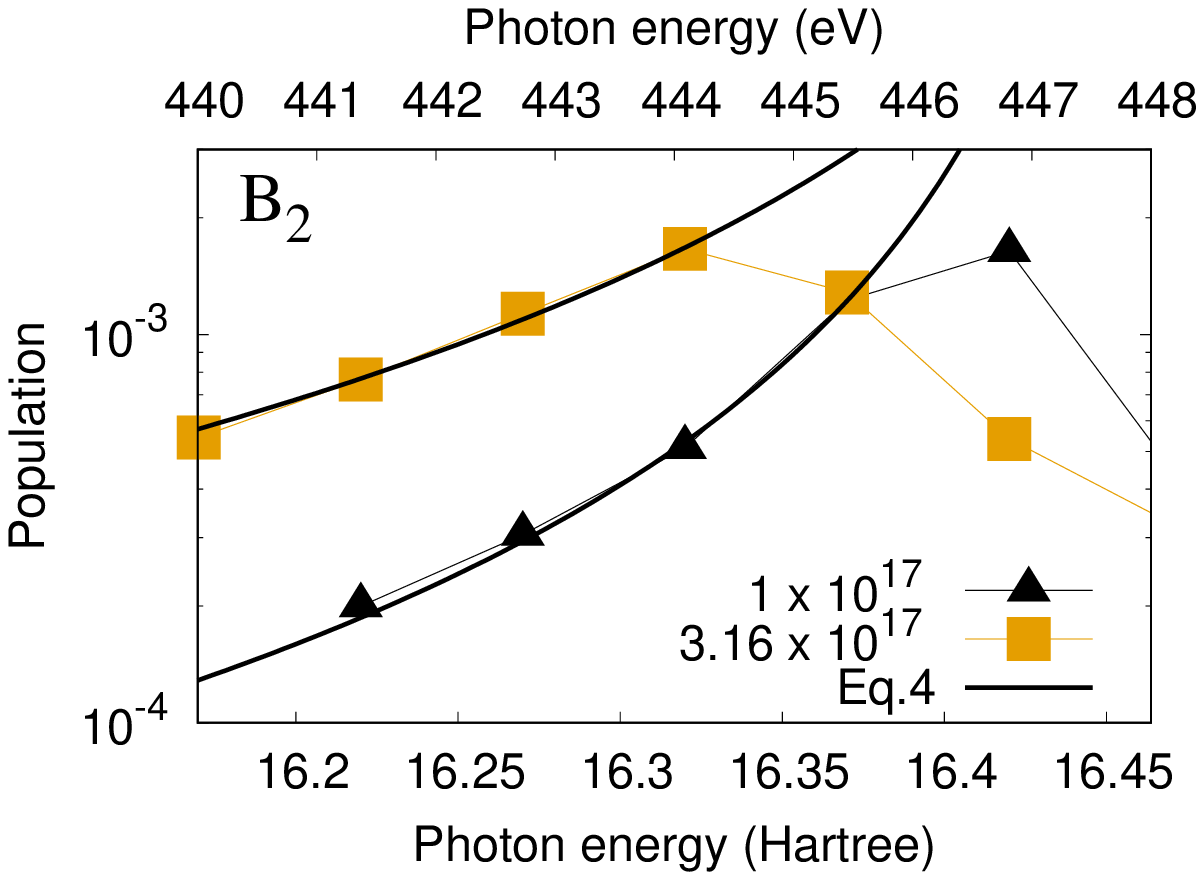}} \\
\end{tabular}
\caption{(Color online) Comparison between population transfer calculated with MCTDHF below the
Oxygen K-edge, and the three-state second-order formula of Eq.~\ref{conteq}, for pulses with intensities
1$\times$10$^{17}$ and 3.16$\times$10$^{17}$ W cm$^{-2}$, as labeled.
The fitted intermediate
state transition energies are listed in Table~\ref{fittable}.
\label{contfig}}
\end{figure*}

The two maxima for population transfer the A$_2$ state, which occurred at 447 and 458eV for low intensity in Fig.~\ref{lowfig},
remain separate as the intensity is increased in Fig.~\ref{popfig}, ending up at approximately 445 and 453eV for intensities
3.16$\times$10$^{17}$ W cm$^{-2}$ and 1.78$\times$10$^{17}$ W cm$^{-2}$, respectively.  The shifts in the maxima for population
transfer from 2nd order to the global optima are 2 and 5eV, respectively.

The results in the prior section demonstrated significant higher-order effects at less than 10$^{16}$ W cm$^{-2}$.
The results in this section make clear that even at modest intensity, 10$^{17}$ W cm$^{-2}$,
strong higher-order effects dominate this nominally second-order process.
The shifting of the maxima to lower photon energies at higher intensities is probably one of the most straightforward signatures
of higher-order effects, being a combination of a 2nd-order AC stark shift with a 2nd-order Raman transition strength for overall
fourth-order behavior.
The development of the minimum in the
valence B$_1$ population transfer is likely driven by a different above-second-order effect, the direct photoionization of the intermediate
core-excited B$_1$ state which leads to efficient depletion if the pulse is resonant, an overall third-order effect.

The main result is that for all three states, 
the optimum population transfer is obtained to the $B_1$ state 
with a photon energy significantly 6eV red-detuned from
the second-order optimum, 0.70\% for the orientation average and 2.39\% for an oriented molecule.
These optima for population transfer with these 1fs linearly polarized pulses
occur at the intensity
3.16$\times$10$^{17}$ W cm$^{-2}$, and a central frequency of 444eV in our calculations, 6eV red-detuned
from the 2nd order optimum at 450eV.

\begin{center}
\begin{table}
\begin{tabular}{|c|c|c|}
\hline
State & Intensity (W cm$^{-2}$) & Intermediate State Energy (eV) \\
\hline
B$_1$ & 1 $\times$ 10$^{17}$ &    449.28  \\
B$_1$ & 3.16 $\times$ 10$^{17}$ &  449.50  \\
\hline
A$_2$ & 1 $\times$ 10$^{17}$ &   446.73  \\
A$_2$ & 3.16 $\times$ 10$^{17}$ & 446.07  \\
\hline
B$_2$ & 1 $\times$ 10$^{17}$ &  448.07  \\
B$_2$ & 3.16 $\times$ 10$^{17}$ &  449.82  \\
\hline
%
%
\end{tabular}
\caption{Fit of intermediate state energy $\hbar \omega_E$ to three-level second-order perturbation theory expression, Eq.~\ref{conteq},
from the fit depicted in Fig.~\ref{contfig}.
\label{fittable}
}
\end{table}
\end{center}

\section{Mechanism of population transfer below edge \label{mechsect}}

The population transfer near the global optimum appears to be
driven by nonresonant Raman transitions.  At the global optimum population transfer -- about 0.8\% to the 
B$_1$ state, at the intensity
3.16$\times$10$^{17}$ W cm$^{-2}$, and a central frequency of 444eV -- the central frequency
of the pulse and the bulk of its 3.25eV bandwidth is substantially red-detuned from the Oxygen K-edge
oscillator strength, the near-edge fine structure and the continuum oscillator strength above the edge.
In Fig.~\ref{popfig}, for each of the three states, the excitation
probability drops steeply as the photon energy is decreased going towards the left side of the figures.  
This strongly decreasing behavior is
consistent with nonresonant Raman due to the Oxygen K-edge oscillator strength.
 In contrast, if population transfer 
were occurring via direct, resonant, one-electron Raman transitions --
via the continuum excitations of the 2$s$ or 2$p$ electron(s) or the 1$s$ electrons
of Nitrogen, and not involving the Oxygen 1$s$ -- we would expect relatively constant behavior on the left side of the figure.

There is much more integrated oscillator
strength available via Oxygen 1s excitations to the the continuum than there is via those to the 
near-edge fine structure, and therefore
nonresonant transitions that are substantially detuned from both the continuum and near-edge fine structure
are likely to proceed via the continuum.  

However, an analysis of these results for 1fs pulses indicates that the mechanism of population
the global optima for population transfer, 
the data are consistent with a nonresonant Raman transition due to the near-edge fine structure, not the continua.
In Fig.~\ref{contfig}, we compare the cross section calculated
for central frequencies 
below the Oxygen K-edge with a simple formula based on second-order perturbation theory.  


Assuming that the transition is driven by one intermediate electronic state, with transition energy $\hbar \omega_E$ from
the ground state, the second-order behavior of the 
population transfer should go approximately as
\begin{equation}
P(\omega_0) \sim \int d\omega \ 
\frac{\vert \mathscr{E}(\omega; \omega_0)\vert^2 \vert \mathscr{E}(\omega - \Delta; \omega_0)\vert^2}{(\omega-\omega_E)^2}
\label{conteq}
\end{equation}
in which expression  $\Delta$ is the excitation energy corresponding
to the Raman transition, and $\mathscr{E}(\omega)$ is the Fourier transfer of the pulse, depending upon central frequency $\omega_0$.
For a narrow bandwidth, or for large detuning ($\omega_0 << \omega_{IE}$), $P(\omega_0)$ goes as $\frac{1}{\omega_0^2}$.
We approximate
the 3.25eV FWHM pulse by a Gaussian, and the Raman transition energy $\Delta$ by the value 3eV.

In Fig.~\ref{contfig}, we plot the population transfer results for
population transfer below the Oxygen K-edge, and the result of fitting the computed values to Eq.~\ref{contfig}.  We extract
an apparent intermediate state transition energy $\hbar\omega_E$ and plot the results in Table~\ref{fittable}.
  The apparent transition energies in Table~\ref{fittable} are near to the positions of the
peaks comprising the near-edge fine structure in Fig.~\ref{xsectfig}.  Also, the difference between the positions at 
1 $\times$ 10$^{17}$ W cm$^{-2}$ and 3.16 $\times$ 10$^{17}$ W cm$^{-2}$ are not very large, a fraction of an eV for
B$_1$ and A$_2$ and 1.75eV for B$_2$, which would seem to rule out an interpretation relying on AC stark shifts of 
the continuum oscillator strength.
The results therefore support the interpretation that the 
Raman population transfer near its global optimum for these 1fs linearly polarized pulses is driven by non-resonant electronic
Raman transitions via the oscillator strength due to the near-edge fine structure.

\section{Conclusions}
 
There is good motivation to obtain more complete predictions to guide the development of
multidimensional
attosecond electronic x-ray Raman spectroscopies~\cite{biggs_two-dimensional_2012, mukamel2013}.
Considerable effort is being directed currently towards the realization of these methods
in the laboratory.  X-ray pulses of the required
coherence, synchronization, and intensity will soon be available with developments in next-generation
light sources or high harmonic generation. 
It is desirable
to provide predictions for the expected efficiencies of these nonlinear methods, which are questionable
due to the major direct loss channels (single and multiple ionization) that are unavoidably present.  The analogy
between these methods and multidimensional nuclear magnetic resonance spectroscopies is tenuous,
because the coupling among electrons is so strong.

The theoretical description of the nonlinear quantum dynamics driven in these proposed
multidimensional spectroscopic methods requires a coherent representation with many active electrons.
Most theoretical and computational treatments so far have not considered the continuum oscillator strength
that may alternatively drive the impulsive population transfer process or provide a loss mechanism, 
and many treatments have been perturbative, explicitly computing only the nth-order
response.  Calculations that do not consider higher-order effects are incapable of predicting the 
conditions required to maximize the amplitude of coherent valence excitations created using 
impulsive electronic Raman transitions in the laboratory.

The MCTHDF method makes it possible to study quantum mechanically coherent nonlinear processes 
at the limits of intensity without making 
any assumptions about the degree of excitation, ionization, or correlation of the wave function.
Using an implementation of the MCTDHF method that is capable of calculating an accurate
solution to the Schrodinger equation including all nonrelativistic electronic effects for polyatomic molecules
using current supercomputer technology,
we have provided a survey of population transfer to valence electronic states in NO$_2$ 
by linearly polarized 1fs pulses tuned near the Oxygen K-edge.  
The results here for fixed nuclei, averaged over orientations, are expected to closely correspond with 
the full result including dynamical nuclear motion, due to the short pulse duration and the absence of 
very light nuclei.

There are omissions in these results that are 
more questionable than the omission of nuclear motion.
Relativistic (broadly speaking, non-dipole) effects are expected to become significant at the
highest intensities, and future work will seek to include these effects using a numerically suitable electromagnetic 
gauge such 
as that described in Ref. CITE.  
%
%
Furthermore, we have not provided any survey of possible pulse shapes or duration.
In future work, we will consider the effect of chirp and duration on the attosecond electronic stimulated Raman 
population transfer process.

However, we have confirmed the convergence 
to within ten percent precision for most aspects of
these results for
population transfer via simulated impulsive electronic x-ray Raman transitions
using 1fs linearly polarized pulses, besides those pertaining to the fixed-nuclei
nonrelativistic approximations, 
including the aspects of the one and many-electron representations, and gauge invariance.

%

%

Our MCTDHF results indicate that
significant (0.70\%) population
transfer to the lowest $B_1$ valence electronic state of the NO$_2$ molecule may be driven
using 1fs x-ray pulses tuned near the Oxygen K-edge. 
This global optimum for maximum population
transfer, orientation-averaged, was found to occur
6eV red-detuned from the 2nd order optimum and 6eV red-detuned from any near-edge fine structure, 
at 3.16$\times$ 10$^{17}$
W cm$^{-2}$.  Preliminary analysis indicated that it proceeds 
via nonresonant Raman transitions through the intermediate core-excited metastable states
comprising the near-edge fine structure, although more work is required to confirm this explanation.

For an oriented molecule, as the molecule will be for subsequent steps of a multidimensional
spectroscopy experiment, the 2.39\% population transfer maximum, greatest among the orientations that we calculated,
 implies a coherence as great as 15\%.  This degree of coherence
would easily enable proposed multidimensional attosecond x-ray spectroscopy experiments.  The cleanest experiment
would be performed at a lesser intensity, to reduce noise.  These calculations indicate that there is no fundamental limitation
to the implementation of proposed multidimensional attosecond x-ray spectroscopy methods, based on the efficiency of
the population transfer with 1fs pulses.


The main result, optimum population transfer for intense pulses substantially red-detuned below edge,
may hold more generally:
multidimensional attosecond electronic X-ray Raman spectroscopies might
in general most efficiently be
performed using intense pulses well red-detuned from resonant edges.  Such pulses may minimize loss through direct and sequential
ionization and make use of the coherent combination of discrete and continuum edge oscillator strength, thereby
providing the greatest potential for creating localized coherent valence electronic wave packets
through impulsive stimulated x-ray Raman excitation.  
Strong red-detuning may provide a way to efficiently perform stimulated
Raman transitions, because it prevents an excursion of the 1$s$ electron.  

Although the simple fit performed in Sec.~\ref{mechsect} seemed to indicate that the transition
at the global optimum for population transfer is driven by discrete excitation, further work will seek to
more fully understand the role of continuum oscillator strength in the population transfer mechanism
both at high intensity, near the global optimum, and at lowest order.  

Owning the wave function, we may test various hypotheses
about the mechanism; MCTDHF allows the problem to be studied without making assumptions about
the mechanism beforehand, in a gauge-invariant manner derived from first principles.
The MCTDHF method, with its unparalleled variational flexibility and first-principles foundation, 
may provide answers to questions involving correlated electronic dynamics, 
and guidance for both models and experiments of the future.
Coupled with recent advancements in 
the greater MCTDHF/MCTDHB/MCTDH effort, such as
the entropy-minimzation techniques [CITE] that may improve the convergence,
or systems of coupled Lindblad equations [CITE] that may allow the determination of many observables
using a small computational domain even with multiple ionization,
will increase its power and utility.  

It remains to be seen whether the more accurate picture of
the nonlinear quantum dynamics driving attosecond stimulated x-ray Raman transitions
near their global optimum 
is that of an excursion of an electron from the 1s shell to the valence space
and back, as the Raman process has been modeled so far in terms of solely discrete excitations,
or whether the more accurate picture is that of a strongly driven, red-detuned 1$s^2$ oscillator, vibrating with both continuum 
and discrete contributions, that drives the valence transition through electron-electron interaction and Fermi repulsion.
In question is whether and how the continuum oscillator
strength may combine coherently to drive the population transfer, at lowest order but also as a function of intensity.  
In question is the
range of the excursion of the 1$s$ electron, at lowest order but also as a function of intensity.  
A greater understanding of the strongly-driven quantum many-body physics that most efficiently drives
 stimulated x-ray Raman transitions 
would inform the development of theories and methods specific to multidimensional attosecond x-ray spectroscopy.

%

\section{Acknowledgments}

This work was supported by the US Department of 
Energy Office of Science, Basic Energy Sciences (BES) and 
Advanced Scientific Computing Research (ASCR) programs,
through contracts DE-AC02-05CH11231 and XXXXXXXXXXX,
primarily through the US DOE Early Career program.
Calculations have been performed on the Lawrencium supercluster at LBNL,
administered by the Laboratory Research Computing center, and
on the machines of the National Energy Research Scientific Computing Center (NERSC), supported
through contracts DE-AC02-05CH11231 and  XXXXXXXXXXXXXXXXX.
We thank our collaborators in the BigStick project for sharing a large amount computer
time, we regret taking so much of it, and we are indebted to
the ASCR Leadership Computing Challenge program from which
it originated.  Further financial support was provided by the Peder Sather Grant program.

\bibliography{no2bib}

\end{document}